\documentclass[journal]{IEEEtran}
\usepackage[explicit]{titlesec}
\usepackage{float} 
\usepackage[figuresright]{rotating}
\usepackage{graphicx}
\usepackage{verbatim}
\usepackage{epstopdf}
\usepackage{subcaption}
\usepackage[justification=justified,singlelinecheck=false]{caption}
\usepackage{algorithmic}
\usepackage{amsmath}
\usepackage{amsthm}
\usepackage{mathrsfs}
\usepackage{extarrows}
\usepackage{cite}

\usepackage{bbm}
\usepackage{amssymb}
\usepackage{booktabs}

\newcommand{\RNum}[1]{\uppercase\expandafter{\romannumeral #1\relax}}

\usepackage{stfloats}

\usepackage{color, soul}
\usepackage{tikz,xcolor,hyperref}
\hypersetup{colorlinks=true, linkcolor=blue, citecolor=blue, urlcolor=blue,}
\usepackage{multirow}
\usepackage{makecell}

\definecolor{lime}{HTML}{A6CE39}

\makeatletter
\DeclareRobustCommand{\orcidicon}{%
	\begin{tikzpicture}
	\draw[lime, fill=lime] (0,0) 
	circle [radius=0.16] 
	node[white] {{\fontfamily{qag}\selectfont \tiny ID}};    \draw[white, fill=white] (-0.0625,0.095) 
	circle [radius=0.007];    \end{tikzpicture}
	\hspace{-2mm}}
\foreach \x in {A, ..., Z}{%
	\expandafter\xdef\csname orcid\x\endcsname{\noexpand\href{https://orcid.org/\csname orcidauthor\x\endcsname}{\noexpand\orcidicon}}
}

\newcommand*\bigcdot{\mathpalette\bigcdot@{.5}}
\newcommand*\bigcdot@[2]{\mathbin{\vcenter{\hbox{\scalebox{#2}{$\m@th#1\bullet$}}}}}
\makeatother

\usepackage[linesnumbered,ruled,vlined]{algorithm2e}

\begin{document}
\title{A Robust Cooperative Vehicle Coordination Framework for Intersection Crossing}
\author{Haojie~Bai, Jiping~Luo,
	 Huafu~Li, Xiongwei~Zhao and~Yang~Wang
\thanks{This work has been supported in part by the Science and Technology Project of Shenzhen under Grant JCYJ20200109113424990, and the Marine Economy Development Project of Guangdong Province under Grant GDNRC [2020]014. (Corresponding author: \textit{Yang Wang}.)
	
Haojie Bai, Xiongwei Zhao and Yang Wang are with the School of Electronic and Information Engineering, Harbin Institute of Technology (Shenzhen), Shenzhen 518071, China (e-mail: hjbai@stu.hit.edu.cn, yangw@hit.edu.cn).

Jiping Luo is with the Department of Computer and Information Science, Link\"oping University, Link\"oping 58183, Sweden (e-mail: jiping.luo@liu.se).
	
Huafu Li is with the China Mobile Information Technology Co., Ltd., Shenzhen 518000, China (e-mail: lihuafu@chinamobile.com).

}}

\maketitle
\allowdisplaybreaks
\begin{abstract}
Cooperative vehicle coordination at unsignalized intersections has garnered significant interest from both academia and industry in recent years, highlighting its notable advantages in improving traffic throughput and fuel efficiency. However, most existing studies oversimplify the coordination system, assuming accurate vehicle state information and ideal state update process. The oversights pose driving risks in the presence of state uncertainty and communication constraint. To address this gap, we propose a robust and comprehensive intersection coordination framework consisting of a robust cooperative trajectory planner and a context-aware status update scheduler. The trajectory planner directly controls the evolution of the trajectory distributions during frequent vehicle interactions, thereby offering probabilistic safety guarantees. 
To further align with coordination safety in practical bandwidth-limited conditions, we propose a context-aware status update scheduler that dynamically prioritizes the state updating order of vehicles based on their driving urgency.
Simulation results validate the robustness and effectiveness of the proposed coordination framework, showing that the collision probability can be significantly reduced while maintaining comparable coordination efficiency to state-of-the-art strategies. Moreover, our proposed framework demonstrates superior effectiveness in utilizing wireless resources in practical uncertain and bandwidth-limited conditions.


\end{abstract}

\IEEEpeerreviewmaketitle

\section{Introduction}\label{sectionI}
Recent advancements in information and control technologies have shown significant potential to enhance the performance of connected and autonomous vehicles (CAVs)\cite{garcia2021tutorial}. Unlike standalone autonomous driving solutions, CAVs share information via vehicle-to-everything (V2X) communication links and make decisions collaboratively to achieve a common goal. This \emph{collectivism} has demonstrated its superiority in driving safety and traffic efficiency\cite{rios2016survey, luo2023computationally}. In recent years, vehicle coordination at critical areas, especially road intersections, has gained substantial research interest and is considered a key enabler for intelligent transportation systems (ITS)\cite{khayatian2020survey}.

Vehicle coordination at unsignalized intersections can be formulated as a multi-agent cooperative trajectory planning problem subject to traffic rules and safety constraints. 
Existing works can be categorized into the rule-based, learning-based, and optimization-based methods\cite{namazi2019intelligent}. Rule-based methods predefine a set of hard-coded and heuristic rules, which are easy to implement and execute quickly \cite{mitrovic2019combined,meng2017analysis}. However, they often suffer from poor coordination efficiency. Learning-based methods emerge as a feasible paradigm that learn complicated coordination strategies through sustained training and fine-tuning \cite{luo2023real,guan2020centralized,liu2024cooperative}. Despite their adaptability in dynamic environments, they rely heavily on pre-collected datasets and function as black boxes with limited interpretability, raising concerns about generalization and safety~\cite{namazi2019intelligent}. 
Optimization-based methods treat the coordination problem as a mathematical program and solve it using well-known tools and algorithms from optimization theory \cite{pei2019cooperative,luo2023computationally,zhang2021trajectory,donglin2024TITS,rios2016automated}. To reduce the complexity of coordination problems, most existing works assume accurate vehicle state information and ideal state update processes.
However, this is not always the case~\cite{khayatian2020survey, dey2015review}. In practice, vehicles face various sources of uncertainty, such as unavoidable noise in onboard sensors (e.g., GPS and inertial measurements)\cite{bai2022time, li2018high} and disturbances in the controller and actuator. Additionally, coordination may lack the freshest information and take risky actions due to communication constraints, such as bandwidth limits and packet drops. Therefore, \emph{ it is crucial to incorporate system uncertainty and imperfection into the algorithm design and propose a robust coordination framework.}


However, only a few works have developed robust coordination strategies. Chalaki \emph{et al.} \cite{chalaki2021priority} introduce a priority-aware resequencing mechanism to handle trajectory disturbances and deviations. Taking into account location uncertainties and prediction errors, Vitale \emph{et al.} \cite{vitale2022autonomous,vitale2022optimizing} employ an open-loop linear predictor to propagate the estimated states in the coordination process. The authors in \cite{pan2023hierarchical} developed a robust coordination strategy to account for model and sensor noises, which applies tube-based model predictive control to solve velocity trajectories based on the robust invariant set. In \cite{chen2022re}, a re-planning strategy is proposed to incorporate accumulated trajectory deviations, thereby enhancing the robustness of predicted instructions. However, these works do not directly handle the evolution of trajectory uncertainties\cite{chalaki2021priority,vitale2022autonomous,vitale2022optimizing,chen2022re}, resulting in potential collision risks in practical driving scenarios. Moreover, to simplify collision region analysis under trajectory uncertainties, the CAVs are either prohibited from turning maneuvers~\cite{vitale2022autonomous, vitale2022optimizing, pan2023hierarchical} or restricted to predefined paths~\cite{chalaki2021priority, vitale2022autonomous, vitale2022optimizing, pan2023hierarchical}. 
The recently developed theory of covariance steering (CS) considers the evolution of state uncertainty and drives it to prescribed targets \cite{okamoto2018optimal}, showing appealing safety guarantees. However, the relevant works primarily focus on single-agent systems and have not yet effectively addressed the issue of interaction among multiple agents.

Furthermore, most aforementioned works overlook the underlying status updating process. However, status update is an integral component of coordination systems and contributes to the evolution of the state and its uncertainty, thereby affecting vehicle maneuvering, especially in bandwidth-limited scenarios~\cite{zheng2020urgency}. Therefore, it is desired to prioritize the state information flow according to the \emph{driving context}~\cite{li2024toward, luo2025TCOM, zheng2024c,  wang2023inp, luo2025TIT}, e.g., critical driving conditions and potential surrounding risks. Intuitively, vehicles in urgent situations (e.g., dense traffic and high collision risks) should be granted priority access to channels for updates. Recent works in \cite{nazari2018remote, nazari2018impact, chohan2019robust,bai2023context} have made strides towards the direction of intersection crossing with
unreliable wireless links. However, these studies focus on analyzing the effects of packet loss and communication frequency, rather than on the design of the underlying scheme.



In outline, robust intersection coordination requires a more comprehensive approach to managing vehicle state uncertainty. This involves directly handling inherent state uncertainties on the one hand, and optimizing the state update process on the other. By simultaneously aligning with coordination efforts, the coordination system is capable of facilitating more robust and critical control for the most urgent vehicles. Thus, further comprehensive studies are required to achieve the above goal in practical driving scenarios.



Motivated by the gap, the main contributions are as follows:
\begin{itemize}
    \item We propose a robust intersection coordination framework that regulates the evolution of vehicle state uncertainty. The framework integrates a robust cooperative trajectory planner and a context-aware status update scheduler through receding horizon optimization and flexible update index, which are designed to plan robust trajectories and assign vehicle status update priorities.
    \item We propose a multi-vehicle receding horizon trajectory planner to direct control the evolution of the state covariance during coordination. To address the frequent interaction of stochastic vehicle states, the collision avoidance chance constraint is analytically formulated in the form of the coupled state mean and covariance of vehicles, which directly control the evolution of the position distribution, thus avoiding potential collisions.
    \item  We propose a context-aware updating scheme that adaptively prioritizes the vehicle state updating order based on the \emph{driving context}. An explicit and flexible update index is derived by holistically assessing critical state conditions and potential surrounding risks, which prioritizes timely updates for the most urgent vehicles to reduce collision risk.
    \item Simulation results demonstrate that our framework not only significantly enhances coordination safety but also improves the effectiveness of communication resource utilization. We further highlight the impact of varying levels of measurement noise, failure probabilities, and transmission success probabilities on coordination safety.
\end{itemize}

The rest of the paper is organized as follows. Section~\ref{section II} describes the intersection model and the vehicle model. Section~\ref{section III} presents the coordination algorithm, where the robust trajectory planner and the context-aware status update scheduler are presented in Section~\ref{section IIIA} and Section~\ref{sectionIIIB}. Simulation results and conclusion are presented in Section~\ref{sectionV} and Section~\ref{sectionVI}.

\begin{table}[h]
	\begin{centering}
		\caption[Notation used in this chapter]{Notations used in this paper.\label{Notation}}
		\par\end{centering}
	\noindent\resizebox{1.0\columnwidth}{!}{%
		\begin{centering}
			\begin{tabular}{|>{\centering}m{0.269\columnwidth}|>{\raggedright}m{0.89\columnwidth}|}
				\hline 
				\textbf{Symbol} & \textbf{\qquad \qquad \qquad \qquad \qquad \qquad Definition}\tabularnewline
				\hline 
				\hline 
				$\bar{x}$, $\hat{x}$, $\tilde{x}$  &  Mean, estimation, estimation error of state $x$.\tabularnewline				
				\hline 
                    $\Sigma$, $\hat{\Sigma}$, $\tilde{\Sigma}$  & Covariance of state, estimation, estimation error. \tabularnewline
				\hline
                    $(\cdot)_i^t$ & The value of $i$-th vehicle at time $t$. \tabularnewline
				\hline
                    $(\cdot)_i^{k \mid t}$ & The Prediction for time $k$ computed at time $t$. \tabularnewline
				\hline
                    $p_i^t$, $\theta_i^t$, $v_i^t$ & Position, longitudinal heading angle (positive counter-clockwise) and velocity. \tabularnewline
				\hline
                    $\mathcal{X}$ &  State space. \tabularnewline
				\hline
                    $u_i^t$ &  Control input. \tabularnewline
                    \hline
				$a_i^t, \delta_i^t$ & Longitudinal acceleration and steering angle (positive counter-clockwise). \tabularnewline
				\hline
                    $G_i^t $ & Process noise matrix. \tabularnewline
				\hline
                     $C_i^t $, $D_i^t $  & Measurement system matrices. \tabularnewline
				\hline
                    $\mathscr{F}_i^t$  & The filtration represents the prior state and observation from $0$ to $t$. \tabularnewline
				\hline
                    $\mathcal{C}_{i j}$  & Collision condition of vehicle $i$ with respect to vehicle $j$. \tabularnewline
				\hline
                    $\mathcal{C}_u$  & Convex polytope of the control input. \tabularnewline
				\hline
                    $\|x\|$ & Euclidean norm of vector $x$. \tabularnewline
				\hline
				 $\|x\|_Q^2=x^T Q x$  & Weighted squared norm with $x$ and matrix $Q$. \tabularnewline
				\hline
				$\mathbb{R}^{n}$, $\mathbb{R}^{m \times n}$ & Set of $n$-dimensional real vectors and $m$ × $n$ real matrices. \tabularnewline
				\hline
				$\mathbb{N}$  & Set of natural numbers. \tabularnewline
				\hline
				 $\mathbb{I}_{a: b}$  & Set of integers $\{a, a+1,...,b\}$. \tabularnewline
				\hline
                  $\succeq$ &  Positive semidefinite. \tabularnewline
				\hline
                    $\mathbb{E}$[·]  & Expectation operation. \tabularnewline
				\hline
				$\operatorname{Pr}$[·]  & Probability of an event. \tabularnewline
				\hline
				$\mathbb{P}$[·] &  Probability density function. \tabularnewline
				\hline
                    $I_n \in \mathbb{R}^{n}$ & Identity matrix of size $n$. \tabularnewline
				\hline
                    $\operatorname{blkdiag}(X_1,...,X_n)$ &  Block-diagonal matrix with matrices $X_1,...,X_n$. \tabularnewline
				\hline
                    $\mathcal{A}_i$, $\mathcal{B}_i$, $\mathcal{R}_i$, $\mathcal{K}_i$, $\mathcal{H}_i$, $\mathcal{L}_i$ & Block matrices over the entire horizon. \tabularnewline
				\hline				
			\end{tabular}
			\par\end{centering}
	}
\end{table}


\begin{figure}[htbp]
	\centering
 \includegraphics[width=0.9\linewidth]{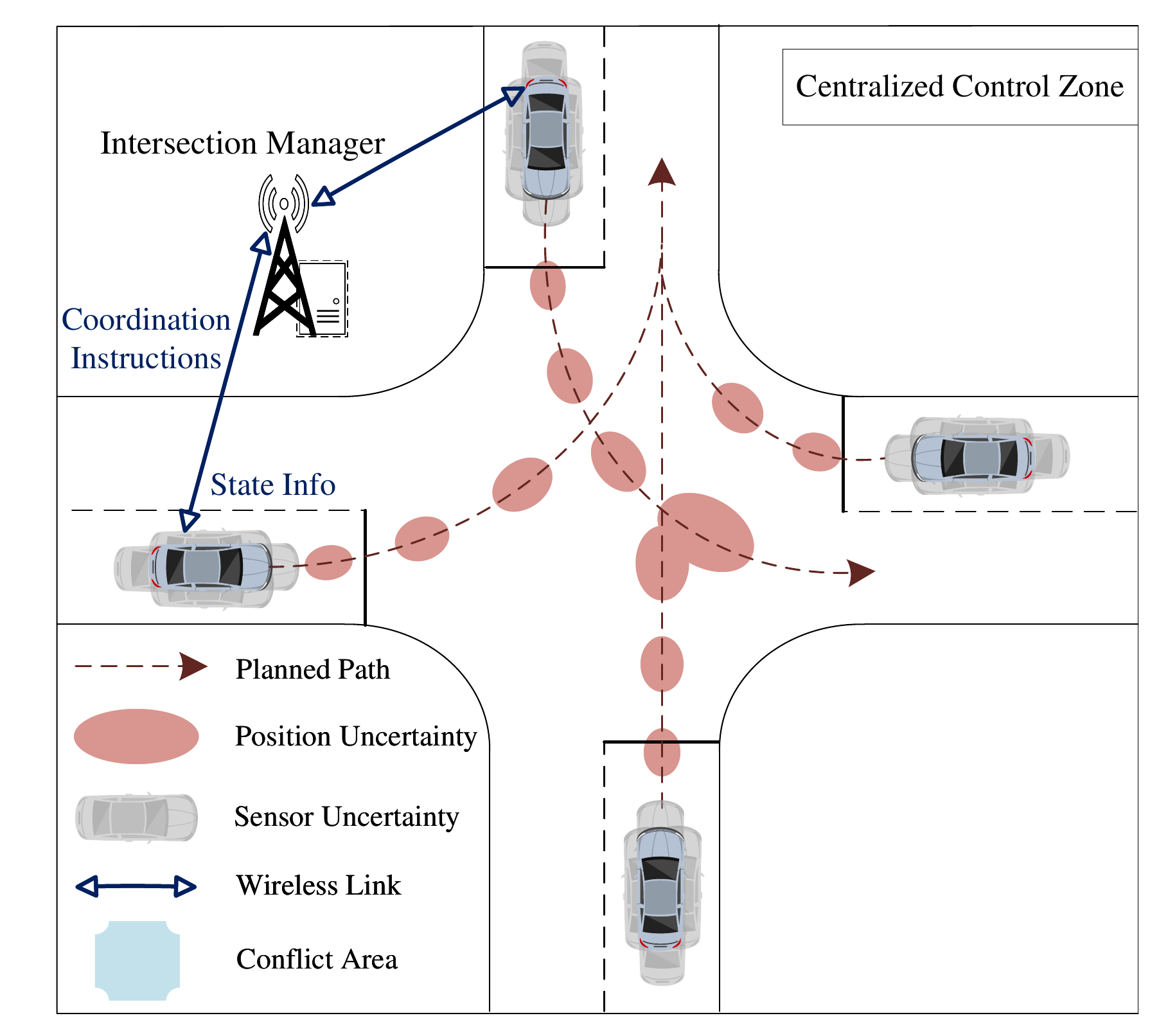}\\
	\caption{Graphic model of intersection coordination system.}\label{Fig01}
	\vspace{-1em}
\end{figure}

\section{System Model}\label{section II}
\subsection{Intersection Coordination Framework}

We consider a typical unsignalized intersection consisting of $4$ roads with lane width $\mathcal{W}_{\text {lane }}$, as depicted in Fig. \ref{Fig01}. The area around the intersection is called the centralized control zone (CCZ) where an intersection manager (IM) is deployed at the multi-access computing (MAC) server to collect the state information of the vehicles and give instructions. The conflict area (CA) represents the central region of the CCZ where multiple paths cross and merge.

IM coordinates approaching CAVs as soon as they are $m$ meters away from the designated intersection. CAVs are not restricted to predefined and fixed paths to improve the utilization of road space and coordination efficiency.
The uncertainty in the coordination maneuvers stems from motion uncertainty, as well as from sensor noise that corrupts the measurements. Only an imprecise approximation of CAVs’ states can be obtained. Unlike most existing intersection coordination methods, which solely rely on vehicle position points for coordination, our approach incorporates state uncertainties and their evolution into the coordination computation. This is aimed at enhancing coordination safety. As illustrated in Fig. \ref{Fig01}, the 3$\sigma$ error of the vehicle position distribution and its corresponding evolution are denoted by the red ellipses. 
 
Each CAV shares its estimated driving states (position, velocity, heading angle, etc.) and intention (go straight, turn right, turn left, etc.) with the IM via vehicle-to-infrastructure (V2I) communication links\footnote{V2I communication is built upon cellular vehicle-to-everything (C-V2X) technology, specifically leveraging the 5G New Radio (NR) cellular network as defined by 3GPP Release 16. The radio interface between the vehicle onboard unit (OBU) and the IM is provided through the Uu interface, which offers robust transmission capabilities (i.e., uplink and downlink), especially for latency-sensitive applications. The V2I link enables the exchange of driving information and coordination instructions between CAVs and IM, forming a closed-loop system and utilizing the centralized architecture and wide coverage of the cellular network \cite{garcia2021tutorial}.}. However, in practical scenarios marked by a scarcity of bandwidth resources, simultaneous transmissions inevitably introduce channel interference and congestion. This could result in packet collisions and drops. During this stage, the state update process affects the evolution of the state and its uncertainty, resulting in additional state uncertainties. Hence, it is essential to adaptively prioritize the state information flow required for coordination.

\begin{figure}
    \centering
    \includegraphics[width=\linewidth]{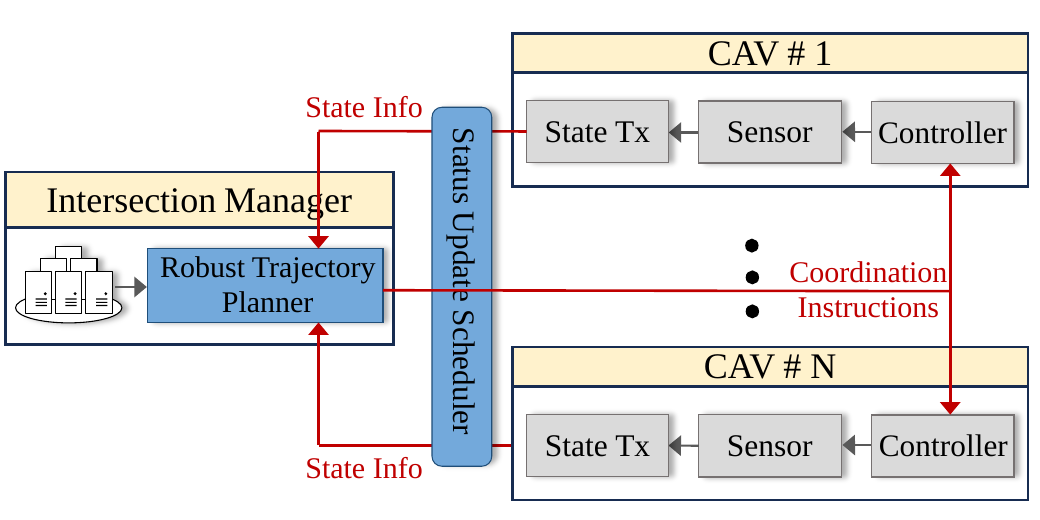}\\
    \caption{Structure of the coordination framework.}\label{Fig02}
    \vspace{-1em}
\end{figure}

To resolve the above barriers, we propose a robust tightly-coupled coordination framework, as shown in Fig. \ref{Fig02}. It consists of two phases: 1) in the trajectory planning phase, the IM utilizes the received states from the updated vehicles along with the predicted states from the non-updated or transmission-failed vehicles, as the starting point. The robust cooperative trajectory planner generates the control instructions for vehicle coordination, which directly control the evolution of vehicle state uncertainties during coordination to enhance driving safety;
and 2) in the status transmission phase, the context-aware update scheduler prioritizes the state updating order of vehicles based on the vehicle’s driving context. The vehicle’s driving context, which includes critical state conditions and potential surrounding risks, is used to assess the vehicle driving urgency. This adaptation prioritizes timely state updates for the most urgent vehicles to reduce driving risks. This process is performed successively as shown in Fig.~\ref{Fig02}. It interleaves trajectory planning and update scheduling through the receding horizon optimization and derived flexible update index, which are discussed in the following Section~\ref{section III}.

\subsection{Vehicle Model}

In this work, we consider a set of vehicles $\mathcal{V}=\left\{1,2, \ldots, N\right\}$ moving on the road, each CAV $i \in \mathcal{V}$ is subject to the stochastic, discrete-time and nonlinear dynamics 
\begin{equation}\label{nonlinear}
x_i^{t+1}=F\left(x_i^t, u_i^t\right) + G_i^t w_i^t,
\end{equation}
for $t \in \mathbb{N}$ denotes the discrete time for $t=0, \ldots, T$, $x_i^t \in \mathbb{R}^{n_x}, u_i^t \in \mathbb{R}^{n_u}$ and $F: \mathbb{R}^{n_x \times n_u} \rightarrow \mathbb{R}^{n_x}$ are the state, control input and nonlinear transition function of the $i$-th CAV, $w_i^t \in \mathbb{R}^{n_x}$ is the i.i.d. standard Gaussian process noise and $G_i^t \in \mathbb{R}^{n_x \times n_x}$ is the process noise matrix, which account for motion uncertainty such as external disturbances and model inaccuracies. $G_i^t$ is defined to characterize the longitudinal and lateral uncertainty levels of the vehicle. As the vehicle orientation changes during movement, $G_i^t$ is adjusted accordingly to reflect variations in uncertainty propagation along the current trajectory. This is achieved by applying the transformation $R(\theta) G_i^t R^{T}(\theta)$, where $\theta$ denotes the vehicle's heading angle and $R(\theta)=\Big[\begin{array}{cc}\cos \theta & -\sin \theta \\ \sin \theta & \cos \theta\end{array}\Big]$ is the rotation matrix.
And $x_i^t \in \mathcal{X}_i$ where $\mathcal{X}_i$ is state space. Furthermore, the position of the CAV in 2D space can be extracted with $p_i^t=\Gamma x_i^t$, where $\Gamma \in \mathbb{R}^{2 \times n_x}$ is defined accordingly.
The vehicle positions are implicitly constrained within the road boundaries by the reference trajectories. Furthermore, to address potential out-of-bounds trajectories, we explicitly specify the feasible state space through road boundaries, which consists of two kinds of convex subsets: 1) strip-shaped convex spaces along the road and 2) matrix-shaped convex spaces located in the center of the intersection. 

\begin{figure}
	\centering
	\includegraphics[width=0.65\linewidth]{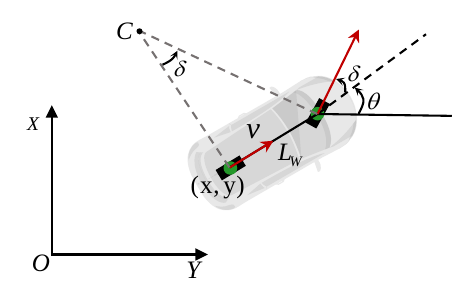}\\
	\vspace{-0.5em}
	\caption{The kinematic model of a front steering vehicle.}\label{Fig03}
	\vspace{-1em}
\end{figure}

As shown in Fig. \ref{Fig03}, the deterministic system dynamics $F\left(x_i^t, u_i^t\right)$ of the vehicle follows the popular bicycle kinematic model, which is widely used for motion modeling in planning and control studies \cite{paden2016survey}.
The bicycle model can effectively approximate the motion of typical front-steered ground vehicles due to its bilateral symmetry and the minimal effects of lateral payload shifts on tire lateral behavior.
The state vector of the $i$-th CAV at time $t$ can be expressed as $x_i^t=\left[\mathrm{x}_i^t, \mathrm{y}_i^t, \theta_i^t, v_i^t\right]^T$, where $\left(\mathrm{x}_i^t, \mathrm{y}_i^t\right)$, $\theta_i^t$ and $v_i^t$ are the coordinates of the rear axle center, longitudinal heading angle (positive counter-clockwise) and velocity, respectively. The control input $u_i^t$ is defined as $\left[a_i^t, \delta_i^t\right]^T$, where $a_i^t$ is the longitudinal acceleration related to the throttle and brake and $\delta_i^t$ is the steering angle (positive counter-clockwise), in this case $n_x=4, n_u=2$. Therefore, the deterministic system dynamics is given as:
\begin{equation}
F\left(x_i^t, u_i^t\right)=\left[\begin{array}{c}
\mathrm{x}_i^t+\tau v_i^t \cos \left(\theta_i^t\right) \\
\mathrm{y}_i^t+\tau v_i^t \sin \left(\theta_i^t\right) \\
\theta_i^t+\tau \frac{v_i^t}{L_w} \tan \left(\delta_i^t\right) \\
v_i^t+\tau a_i^t
\end{array}\right],
\end{equation}
where $\tau$ is the discrete-time interval and $L_w$ is the vehicle wheelbase.  In addition, each CAV sends the estimated state to the IM. The state is observed through the measurement process
\begin{equation}\label{measurement}
z_i^t=C_i^t x_i^t+D_i^t \nu_i^t,
\end{equation}
where $z_i^t \in \mathbb{R}^{n_z}$ is the measurement of the vehicle and $\nu_i^t \in  \mathbb{R}^{n_z}$ is the measurement noise which is the i.i.d. standard Gaussian random vector. $C_i^t \in \mathbb{R}^{n_z \times n_x}$ and $D_i^t \in \mathbb{R}^{n_z \times n_z}$ are the measurement matrices which can be modeled according to the given sensors properties. The matrix $D_i^t$ is assumed to be invertible in order to simplify the filtering equations and $w_i^{0:T}$, $\nu_i^{0:T}$ are independent. We defined the estimated state as $\hat{x}_i^t=\mathbb{E}\left(x_i^t \mid \mathscr{F}_i^t\right)$ and the prior estimated state as $\hat{x}_i^{t^{-}}=\mathbb{E}\left(x_i^t \mid \mathscr{F}_i^{t-1}\right)$, thus the corresponding estimation errors are denoted as $\tilde{x}_i^t=x_i^t-\hat{x}_i^t$ and $\tilde{x}_i^{t^{-}}=x_i^t-\hat{x}_i^{t^{-}}$.

\section{Robust Coordination Framework}\label{section III}


This section presents the details of the intersection coordination framework. The design involves two challenges: 1) The state uncertainty of each CAV consistently increases due to the accumulation of stochastic noise in vehicle movement. If we simply consider the entire timeline, this could result in the convex safe state set becoming too small, eliminating many qualified solutions. Therefore, we need to develop an approach that can simultaneously consider the evolution of these uncertainties and address the coordination among the stochastic and uncertain states of the vehicles. 2) The second challenge lies in the coupling between trajectory planning and status updates. The efficacy (i.e., safety and efficiency) of computed trajectories can only be evaluated given the status updating schemes of the vehicles, meaning that trajectory planning is closely intertwined with status updates.

This section aims to address these challenges. We start our exposition with the cooperative trajectory planner in Section \ref{section IIIA}, and then we discuss the context-aware status update scheduler in Section \ref{sectionIIIB}.


\subsection{Robust Cooperative Trajectory Planner}\label{section IIIA}
The basic idea of the trajectory planner is to coordinate in terms of the first two moments of the vehicle state, that is, mean and covariance, which aims to enhance the safety of frequent multi-vehicle interactions. 
We begin by formulating the cooperative trajectory optimization problem. Then we reformulate the inter-vehicle collision avoidance chance constraint in the form of the coupled state mean and covariance to handle the interaction of stochastic vehicle states. Subsequently, we reformulate the original stochastic problem involving the inaccessible state into the tractable control problem with accessible filtered state.

\textit{1) Problem Formulation:}
Each CAV is considered to follow the stochastic, discrete-time, and nonlinear dynamics Equation \ref{nonlinear} in the coordination. The initial state $x_i^0$ is given by a multivariate Gaussian distribution $x_i^0 \sim \mathcal{N}\left(\mu_i^0, \Sigma_i^0\right)$, where $\mu_i^0 \in \mathbb{R}^{n_x}$ is the mean and $\Sigma_i^0 \in \mathbb{R}^{n_x \times n_x}$ is the covariance matrix, respectively. The states of the vehicles are random variables. To add robustness to the trajectory planner under stochastic uncertainty, we consider the following collision avoidance chance constraint, which enforces that the probability of a collision-free state exceeds a threshold
\begin{equation}\label{avoidc}
\operatorname{Pr}\left(x_i^{k \mid t}\notin \mathcal{C}_{i j}^{k \mid t}\right) \geq 1-\xi_{\text {coll }}^x, \quad i, j \in \mathcal{V}, j \neq i,
\end{equation}
where $\xi_{\text {coll }}^x \in(0,0.5)$ is the pre-specified threshold allowed probability of inter-vehicle collision. The collision condition of vehicle $i$ with respect to vehicle $j$ at time $t$ is denoted as
\begin{equation}
\mathcal{C}_{i j}^{k \mid t}:=\left\{x_i^{k \mid t} \; \Big| \big\|p_i^{k \mid t}-p_j^{k \mid t}\big\| \leq d_{i j}\right\}, k \in \mathbb{I}_{t: t+M},
\end{equation}
where $d_{i j} \in \mathbb{R}$ is the minimum safety distance between the barycenters of vehicles $i$ and $j$. 
In practical scenarios, strictly enforcing hard constraints on control inputs can be overly restrictive or even infeasible due to environmental uncertainties or system noise. To address this, the control input needs to incorporate tolerance for these uncertainties to enhance the robustness and stability of the system. Hence, similar to the state chance constraint, we impose the control input chance constraints of each CAV, i.e.,
\begin{equation}\label{controlc}
\operatorname{Pr}\left(u_i^{k \mid t} \in \mathcal{C}_u\right) \geq 1-\xi_{\text {fail }}^u, \quad i \in \mathcal{V},
\end{equation}
where $\mathcal{C}_u$ is a typical convex polytope.
To address the potential issue of control fluctuations and unsmooth movements, we also impose a direct constraint on the jerk that characterizes the rate of variation of control inputs.
\begin{equation}\label{jerk}
-\kappa_{\max } \leq \kappa_i^{k \mid t} \leq \kappa_{\max }, \quad i \in \mathcal{V}, k \in \mathbb{I}_{t: t+M-1},
\end{equation}
where $\kappa_{\max }$ is the maximum jerk value, which limits the magnitude of the change of control inputs.

At the same time, the objective function of the $i$-th vehicle is
\begin{equation}\label{obf1}
J_i(u_i)=\mathbb{E}\left[ \sum_{k=t}^{t+M}\!\left\|x_i^{{k} \mid t}-x_{i}^{r}\right\|_{Q_i^{k}}^2+\!\sum_{k=t}^{t+M-1}\!\!\left\|u_i^{k \mid t}\right\|_{R_i^k}^2\right],
\end{equation}
with a short prediction horizon $M$, weight matrices $Q_i^k \succeq 0$ and $R_i^k  \succ 0$, and desired state references $x_i^{r}$ sampled from the reference path of each vehicle at the maximum passing speed. Thus, the efficiency of the intersection crossing will be ensured.
Note that the absolute magnitude of the weights does not affect the optimization output; rather, it is the relative magnitude that needs to be adjusted. The weights are determined through a combination of preliminary empirical analysis and fine-tuning experiments. For application scenarios with significant variation, the weight factors need to be adjusted to account for varying constraints, priorities, or operational requirements.

Considering vehicle motion and measurement noises and disturbances, it is crucial to robustly coordinate vehicles to ensure safety and efficiency in intersection management. To achieve this goal, the cooperative trajectory planning problem can be formulated to find the optimal control policies $\boldsymbol{u}$ for all vehicles as follows
\begin{align*} 
\begin{aligned}
\mathscr{P}_1\!: & \min _{\boldsymbol{u}\in \Pi} \mathbb{E}\left[ \sum_{i=1}^N\ \!\!\! \Big(\!\sum_{k=t}^{t+M}\!\left\|x_i^{{k} \mid t}-x_{i}^{r}\right\|_{Q_i^{k}}^2+\!\sum_{k=t}^{t+M-1}\!\!\left\|u_i^{k \mid t}\right\|_{R_i^k}^2\!\Big)\right] \\
\text { s.t. } & x_i^{{k+1} \mid t}=F\left(x_i^{k \mid t}, u_i^{k \mid t},  w_i^{k \mid t}\right), \!\!\! \quad k \in \mathbb{I}_{t: t+M-1} \\
& \operatorname{Pr}\left(x_i^{k \mid t} \notin \mathcal{C}_{i j}^{k \mid t}\right)\! \geq 1-\xi_{\text {coll }}^x, \!\!\!\quad i \neq j, k \in \mathbb{I}_{t: t+M} \\
& \operatorname{Pr}\left(u_i^{k \mid t} \in \mathcal{C}_u\right) \geq 1-\xi_{\text {fail }}^u, \!\!\!\quad i \in \mathcal{V}, k \in \mathbb{I}_{t: t+M-1} \\
&-\kappa_{\max } \leq \kappa_i^{k \mid t} \leq \kappa_{\max }, \quad i \in \mathcal{V}, k \in \mathbb{I}_{t: t+M-1}, \\
& x_i^{k \mid t} \in \mathcal{X}_i.\\
\end{aligned}
\end{align*} 
where $\Pi$  denotes the set of causal policies of the vehicles. Similar to \cite{zheng2024cs}, since the prediction horizon is short and the control input at time $t$ is an affine function of the measurement data, it follows that the states will be a Gaussian distribution over the entire horizon. The $\mathscr{P}_1$ is a non-convex and nonlinear problem for considering nonlinear dynamics and probabilistic constraints in the intersection passing. In the subsequent analysis, we try to solve the robust trajectory optimization problem. The beginning of the $t$-th optimization iteration is at $t$-th time slot. For the sake of clarity, we omit the time reference ``t" from the superscript during the processing.

\textit{2) Safety Constraint Formulation:}
Next, we address the inter-vehicle collision avoidance chance constraint. Given the mean and uncertainty covariances of the two vehicles’ positions $p_i^k \sim \mathcal{N}\left(\bar{p}_i^k, P_i^k\right)$, $p_j^k \sim \mathcal{N}\left(\bar{p}_j^k, P_j^k\right)$. $p_i^k$ and $p_j^k$ are assumed as independent multivariate Gaussian distributions, then $p_i^k-p_j^k$ is also a multivariate Gaussian distribution, i.e. $p_i^k-p_j^k \sim \mathcal{N}\left(\bar{p}_i^k-\bar{p}_j^k, P_i^k+P_j^k\right)$. Therefore, the instantaneous collision probability of vehicle $i$ with vehicle $j$ can be written as a double integral of a multivariate Gaussian probability density function over a circle
\begin{equation}
\operatorname{Pr}\left(x_i^k \in \mathcal{C}_{i j}^k\right)=\int_{\left\|p_i^k-p_j^k\right\| \leq d_{i j}} \mathbb{P}\left(p_i^k-p_j^k\right) d\left(p_i^k-p_j^k\right).
\end{equation}

However, there is no analytical form to express the collision probability, here we introduce a procedure to approximate the circular collision region $\mathcal{C}_{i j}^k$ with a half space $\tilde{\mathcal{C}}_{i j}^k$ around the previous nominal trajectories, which is denoted as
\begin{equation}
\tilde{\mathcal{C}}_{i j}^k:=\left\{x \mid \alpha_{i j}^{k^T}\left(p_i^k-p_j^k\right) \leq d_{i j}\right\},
\end{equation}
where $\alpha_{i j}^k=\left(\bar{p}_i^k-\bar{p}_j^k\right) /\left\|\bar{p}_i^k-\bar{p}_j^k\right\|$. The construction results in the convex approximation $\tilde{\mathcal{C}}_{i j}^k$ of the original non-convex forbidden region $\mathcal{C}_{i j}^k$ by linearization. It is evident that $\mathcal{C}_{i j}^k \subset \tilde{\mathcal{C}}_{i j}^k$, thus, we can obtain an approximated collision probability upper bound $\operatorname{Pr}\left(x_i^k \in \mathcal{C}_{i j}^k\right) \leq \operatorname{Pr}\left(x_i^k \in \tilde{\mathcal{C}}_{i j}^k\right)$. Given that $x$ follows a Gaussian distribution, the original chance constraint can be converted into a deterministic constraint \cite{zhu2019chance}.
Therefore, we can obtain a collision probability upper bound between two vehicles:
\begin{equation}
\operatorname{Pr}\!\left(x_i^k \in \tilde{C}_{i j}^k\right)\!=\frac{1}{2}+\frac{1}{2} \operatorname{erf}\!\Bigg(\!\frac{d_{i j}-\alpha_{i j}^{k^T}\left(\bar{p}_i^k-\bar{p}_j^k\right)}{\sqrt{2 \alpha_{i j}^{k^T}\left(P_i^k+P_j^k\right) \alpha_{i j}^k}}\Bigg) \!\leq \xi_{\text {coll }}^x,
\end{equation}
where $\operatorname{erf}(x)$ is the standard error function. Since $\xi_{\text {coll }}^x \in(0,0.5)$ and $\operatorname{erf}(x)$ is a monotonically increasing function, as a result, the collision avoidance chance constraint in Equation \ref{avoidc} can be converted into a deterministic constraint on the vehicle's state mean and covariance
\begin{equation}\label{avoidc1}
\alpha_{i j}^{k^T}\!\!\left(\bar{p}_i^k-\bar{p}_j^k\right)-d_{i j} \geq \operatorname{erf}^{-1}\!\!\left(1-2 \xi_{\text {coll }}^x\right)\!\! \sqrt{2 \alpha_{i j}^{k^T}\!\left(P_i^k+P_j^k\right)\! \alpha_{i j}^k}.
\end{equation}


\textit{3) Robust Trajectory Control Design:}
For approaching CAVs, the IM performs robust trajectory coordination in a receding horizon fashion. Within each optimization iteration, we start by linearizing the dynamics Equation \ref{nonlinear} around some nominal trajectories. The nominal trajectory for each iteration is the trajectory derived from the control sequence of the previous iteration, apart from the nominal trajectory of the first iteration, which can be an arbitrary trajectory. Let $\bar{X}_i^t=\left[\bar{x}_i^{t}, \ldots, \bar{x}_i^{t+M }\right]^T$ and $\bar{U}_i^t=\left[\bar{u}_i^{t }, \ldots, \bar{u}_i^{t+M-1 }\right]^T$ denote the nominal state and control sequence, respectively, at the $t$-th iteration. This procedure results in the stochastic and linear time-variant dynamics
\begin{equation}\label{original}
x_i^{k+1 }=A_i^{k } x_i^{k }+B_i^{k } u_i^{k }+r_i^{k }+G_i^{k } w_i^{k }, k \in \mathbb{I}_{t: t+M-1},
\end{equation}
where $G_i^t \in \mathbb{R}^{n_x \times n_x}$ is the process noise matrix, which depends on the properties of the given scenario. The system matrices $A_i^{k } \in \mathbb{R}^{n_x \times n_x}$, $B_i^{k } \in \mathbb{R}^{n_x \times n_u}$ and residual term $r_i^{k } \in \mathbb{R}^{n_x}$ are given by
\begin{equation}\label{linear}
A_i^{k }\!=\!\left.\frac{\partial F}{\partial x_i^{k}}\right|_{\substack{x_{i}^{k }= \bar{x}_{i}^{k}, u_i^{k}=\bar{u}_i^{k}}}, \quad
B_i^{k}=\left.\frac{\partial F}{\partial u_i^{k }}\right|_{\substack{x_i^{k}=\bar{x}_i^{k}, u_i^{k}=\bar{u}_i^{k}}},
\end{equation}
\begin{equation}\label{residual}
r_i^{k }=F_i\left(\bar{x}_i^{k }, \bar{u}_i^{k }\right)-A_i^k \bar{x}_i^{k }-B_i^k \bar{u}_i^{k }.
\end{equation}

Since the true states of the vehicles are not available, the Kalman estimator is used to estimate the states. It can be shown that each CAV’s estimation error covariance $\tilde{\Sigma}_i^k$ is deterministic, which is determined by the Kalman filter and does not depend on the control over the entire horizon. Substituting $x_i^k=\hat{x}_i^k+\tilde{x}_i^k$ and using properties of conditional expectation, the estimation error covariance $\tilde{\Sigma}_i^k$ in the objective function can be discarded \cite{ridderhof2020chance}. Thus, the objective function in Equation \ref{obf1} can be rewritten as over the entire horizon
\begin{equation}\label{obf11}
\hat{J}(\boldsymbol{u})= \mathbb{E}\!\left[ \sum_{i=1}^N\ \!\!\! \Big(\!\sum_{k=t}^{t+M}\!\left\|\hat{x}_i^{{k} \mid t}\!\!-x_{i}^{r}\right\|_{Q_i^{k}}^2\!\!+\!\!\!\! \sum_{k=t}^{t+M-1}\!\!\left\|u_i^{k \mid t}\right\|_{R_i^k}^2\!\Big)\right].
\end{equation}
By defining the innovation process $\tilde{z}_i^{k^{-}}$ of each CAV as 
\begin{equation}
\tilde{z}_i^{k^{-}}=z_i^k-\mathbb{E}\left(z_i^k \mid \mathscr{F}_i^{k-1}\right)=C_i^k \tilde{x}_i^{k^{-}}+D_i^k \nu_i^k,
\end{equation}
where $\mathbb{E}\left(z_i^k \mid \mathscr{F}_i^{k-1}\right)=\mathbb{E}\left(C_i^k x_i^k+D_i^k \nu_i^k \mid \mathscr{F}_i^{k-1}\right)=C_i^k \hat{x}_i^{k^{-}}$, the estimated state dynamics of each CAV can be written as
\begin{equation}\label{estimation}
\hat{x}_i^{k+1}=A_i^k \hat{x}_i^k+B_i^k u_i^k+r_i^k + K_i^{k+1} \tilde{z}_i^{(k+1)^{-}}.
\end{equation}
where $K_i^{k+1}$ is the Kalman gain. The state and measurement process Equations \ref{original} and \ref{measurement} are essentially replaced by the corresponding estimated state process Equation \ref{estimation} with the corresponding noise $K_i^{k+1} \tilde{\boldsymbol{z}}_i^{(k+1)^{-}}$. 
Therefore, IM coordinates vehicles in terms of the accessible estimated state. 

We write the estimated state process Equation \ref{estimation} of each CAV over the entire horizon in compact form as
\begin{equation}\label{estiinnovation}
\hat{X}_i=\mathcal{A}_i \hat{x}_i^t+\mathcal{B}_i U_i+\mathcal{R}_i+\mathcal{K}_i \tilde{Z}_i.
\end{equation}
Let $\hat{X}_i$, $\tilde{Z}_i$ and $U_i$ be column vectors augmented by stacking $\hat{x}_i^k$, $\tilde{z}_i^{k^{-}}$ and $u_i^k$, i.e., $\hat{X}_i=\left[\hat{x}_i^t, \ldots, \hat{x}_i^{t+M}\right]^T \in \mathbb{R}^{(M+1) n_x}$, $\tilde{Z}_i=\big[\tilde{z}_i^{t^{-}}, \ldots, \tilde{z}_i^{(t+M)^{-}}\big]^T \in \mathbb{R}^{(M+1) n_z}$ and $U_i=\left[u_i^t, \ldots, u_i^{t+M-1}\right]^T \in \mathbb{R}^{M n_u}$. And the block matrices $\mathcal{A}_i$, $\mathcal{B}_i$, $\mathcal{R}_i$ and $\mathcal{K}_i$ are constructed appropriately.


The IM devises the control law by collecting the states of the vehicles once in each optimization iteration. The states are either transmitted by the vehicles or propagated by the IM, which corresponds to whether vehicles are scheduled for status updates.
In this work, we use affine disturbance feedback control policy initially introduced in \cite{balci2021covariance}. To reduce the computational burden, we truncate the disturbance history feedback policy to a length of one, thus decreasing the number of decision variables by $(M-2)^2/2$. As validated in \cite{balci2021covariance}, the truncation yields a substantial improvement in computational efficiency, with only a slight sacrifice in optimization performance. The control input at each discrete stage can be expressed as follows:
\begin{equation}\label{feedbackforward}
u_i^k\!=\!\left\{\begin{array}{cc}
\!\!\!\!\!\bar{u}_i^t+H_i^t\left(\hat{x}_i^t-\mu_i^t\right) & \!\!\!\!\!\!\!\!\!\!\!\!\!\!\!\!\!\!\!\!\!\!\!\!\!\!\!\!\!\!\!\!\!\!\!\!\!\!\!\!\text { if } k=t, \\
\!\!\!\bar{u}_i^k+H_i^k\left(\hat{x}_i^t-\mu_i^t\right)+L_i^k \tilde{z}_i^{k^{-}} & \!\!\!\text {if } k \in[t\!+\!1, t\!+\!M\!\!-\!1],
\end{array}\right.
\end{equation}
where $\bar{u}_i^k \in \mathbb{R}^{n_u}$ are the feedforward gains, $H_i^k \in \mathbb{R}^{n_u \times n_x}$ and $L_i^k \in \mathbb{R}^{n_u \times n_z}$ are the feedback matrices. We write the decision variables in a more compact form, i.e., $\bar{U}_i=\left[\bar{u}_i^t ; \bar{u}_i^{t+1} ; \ldots ; \bar{u}_i^{t+M-1}\right]$, $L_i=\operatorname{blkdiag}\left(L_i^{t+1}, L_i^{t+2}, \cdots, L_i^{t+M-1}\right)$, $\mathcal{L}_i=\Big[\begin{array}{lll}0 & 0 & 0 \\ 0 & L_i & 0\end{array}\Big]$, and $\mathcal{H}_i=\left[H_i^t ; H_i^{t+1} ; \ldots ; H_i^{t+M-1}\right]$. It follows that $U_i=\bar{U}_i+\mathcal{H}_i\left(\hat{x}_i^t-\mu_i^t\right)+\mathcal{L}_i \tilde{Z}_i$, where the first control input of the current horizon is executed. 
Then the estimated state process Equation \ref{estiinnovation} of each vehicle is thus obtained 
\begin{equation}\label{estmatrix}
\hat{X}_i=\mathcal{A}_i \hat{x}_i^t+\mathcal{B}_i \bar{U}_i+\mathcal{B}_i \mathcal{H}_i\left(\hat{x}_i^t-\mu_i^t\right)+\left(\mathcal{K}_i+\mathcal{B}_i \mathcal{L}_i\right) \tilde{Z}_i+\mathcal{R}_i,
\end{equation}
where $\tilde{Z}_i$ has zero mean since $\tilde{z}_i^{k^{-}}$ is a Gaussian process and the covariance of the innovation process is the block-diagonal matrix since each step of $\big(\tilde{z}_i^{k^{-}}\big)$ is independent \cite{aastrom2012introduction}, i.e., $\Sigma_{\tilde{Z}_i}=\mathbb{E}\big(\tilde{Z}_i \tilde{Z}_i^T\big)=\operatorname{blkdiag}\big(\Sigma_{\tilde{z}_i^{t^{-}}}, \ldots, \Sigma_{\tilde{z}_i^{(t+M)^{-}}}\big)$. Similar to \cite{bai2023context}, \cite{ridderhof2020chance}, the covariance of each innovation process is obtained as
\begin{equation}
\Sigma_{\tilde{z}_i^{k^{-}}}=\mathbb{E}\big(\tilde{z}_i^{k^{-}} \tilde{z}_i^{k^{-T}}\big)=C_i^k \tilde{\Sigma}_i^{k^{-}} C_i^{k^T}+D_i^k D_i^{k^T}.
\end{equation}

Considering whether to update the status at the beginning of the current iteration, then we can obtain the mean by taking the expectation of both sides of Equation \ref{estmatrix} and compute the covariance of the estimated state process as follows:
\begin{equation}\label{X_mean}
\bar{X}_i=\mathbb{E}\big(\hat{X}_i\big)=V_i^t S_i^t\mathcal{A}_i(\hat{x}_i^t-\mu_i^t)+ \mathcal{A}_i \mu_i^t+\mathcal{B}_i \bar{U}_i+\mathcal{R}_i,
\end{equation}
\begin{align} \label{X_cov}
\hat{\Sigma}_i=&\left(1-V_i^t S_i^t\right)\left(\mathcal{A}_i+\mathcal{B}_i \mathcal{H}_i\right) \hat{\Sigma}_i^t\left(\mathcal{A}_i+\mathcal{B}_i \mathcal{H}_i\right)^T \nonumber\\
&+\left(\mathcal{K}_i+\mathcal{B}_i \mathcal{L}_i\right) \Sigma_{\tilde{Z}_i}\!\left(\mathcal{K}_i+\mathcal{B}_i \mathcal{L}_i\right)^T,
\end{align} 
where $\hat{x}_i^t$ and $\hat{\Sigma}_i^t$ denote the initial estimated state and corresponding covariance, respectively in each iteration. Note that the update decision $V_i^t$ could have been obtained through the status update scheduling at the start of each iteration, which will be discussed in detail in Section \ref{sectionIIIB}. For the sake of brevity, we will temporarily omit the known update decision terms in the following expressions. Next we compute the covariance of each control process as
\begin{equation}\label{U_meancov}
\Sigma_i^U=\mathcal{H}_i \hat{\Sigma}_i^t \mathcal{H}_i^T+\mathcal{L}_i \Sigma_{\tilde{Z}_i} \mathcal{L}_i^T.
\end{equation}
By substituting $\hat{X}_i= \hat{X}_i - \bar{X}_i +\bar{X}_i$ and $U_i= U_i - \bar{U}_i +\bar{U}_i$, and leveraging the properties of the trace operator, the cost function in Equation \ref{obf11} can be rewritten as follows
\begin{align}
\hat{J}\big(\bar{X}_i, \hat{\Sigma}_i, &\bar{U}_i, \Sigma_i^U\big)=\sum_{i=1}^N\big[ \!\operatorname{tr}\!\big(Q_i \hat{\Sigma}_i +R_i \Sigma_i^U\big) \nonumber \\
&+\big(\bar{X}_i\!-\!X_{i}^{r}\big)^T \!Q_i\big(\bar{X}_i\!-\!X_{i}^{r}\big)\!+\bar{U}_i R_i \bar{U}_i\big],
\end{align}
where $Q_i=\operatorname{blkdiag} \left(Q_i^{t}, Q_i^{t+1}, \cdots Q_i^{t+M}\right) \succeq 0$ and $R_i=\operatorname{blkdiag} \left(R_i^t, R_i^{t+1}, \cdots R_i^{t+M-1}\right) \succ 0$. The cost function is reformulated in terms of the first and second moments of the state, i.e., the mean and covariance, enabling joint optimization over both the magnitude and uncertainty of the state, thereby facilitating more robust control.
Then, substituting the mean and covariance of the state and control input in Equations \ref{X_mean}, \ref{X_cov}, \ref{U_meancov} and combining like terms, one can further convert the cost function into
\begin{align}\label{longequ}
&\hat{J}(\boldsymbol{\bar{U}}, \boldsymbol{\mathcal{H}}, \boldsymbol{\mathcal{L}})\!=\!\!\sum_{i=1}^N \! \Big[\!(\mathcal{A}_i \mu_i^t\!+\mathcal{B}_i \bar{U}_i+\! \mathcal{R}_i \!-\! X_{i}^{r})^T \! Q_i(\mathcal{A}_i \mu_i^t+ \mathcal{B}_i \bar{U}_i \nonumber\\
&+\! \mathcal{R}_i \!-\! X_{i}^{r}) \!+\! \bar{U}_i^T R_i \bar{U}_i \Big] \!\!	+\! \sum_{i=1}^N \! \operatorname{tr}\! \Big\{\! \Big[\! \left(\mathcal{A}_i \!+\! \mathcal{B}_i \mathcal{H}_i\right)^T \!\! Q_i \left(\mathcal{A}_i \!+\!\mathcal{B}_i \mathcal{H}_i\right) \nonumber\\
&+\! \mathcal{H}_i^T \!R_i \mathcal{H}_i\Big] \hat{\Sigma}_i^t \!+\! \left[\left(\mathcal{K}_i \!+\! \mathcal{B}_i \mathcal{L}_i\right)^T \!\! Q_i \! \left(\mathcal{K}_i \!+\! \mathcal{B}_i \mathcal{L}_i\right) \!+\! \mathcal{L}_i^T \!R_i \mathcal{L}_i\right] \! \Sigma_{\tilde{Z}_i} \!\Big\}
\end{align}
which is convex in $\bar{U}_i, \mathcal{H}_i$ and $\mathcal{L}_i$.

Next, using the estimated state, we rewrite the inter-vehicle collision avoidance chance constraint in Equation \ref{avoidc} as
\begin{equation}\label{aviodc11}
\operatorname{Pr}\left(\hat{\mathrm{x}}_i^k+\tilde{x}_i^k \notin C_{i j}^k\right) \geq 1-\xi_{\text {coll }}^x, \quad i, j \in \mathcal{V}, j \neq i.
\end{equation}
Let $E_x^k \triangleq\left[\begin{array}{lll}0_{n_x, k n_x} & \mathrm{I}_{n_x} & 0_{n_x,(M-k) n_x}\end{array}\right]$, and the mean $p_i^k$ and uncertainty $P_i^k$ of the position in Equation \ref{avoidc1} at time $k$ can be written as
\begin{equation}\label{mean}
p_i^k=\Gamma E_x^k\left(\mathcal{A}_i \mu_i^t+\mathcal{B}_i \bar{U}_i+\mathcal{R}_i\right),
\end{equation}
\begin{equation}\label{covariance}
P_i^k=\Gamma E_x^k \hat{\Sigma}_i E_x^{k^T} \Gamma^T+\Gamma \tilde{\Sigma}_i^k \Gamma^T.
\end{equation}
We can then write the inter-vehicle collision avoidance constraints in Equation \ref{aviodc11} by Equations \ref{avoidc1}, \ref{mean} and \ref{covariance}, and obtain the second-order cone constraint
\begin{align}\label{avoidc111}
&\alpha_{i j}^{k^T} \Gamma E_x^k\left(\mathcal{A}_i \mu_i^t+\mathcal{B}_i \bar{U}_i+\mathcal{R}_i-\mathcal{A}_j \mu_j^t+\mathcal{B}_j \bar{U}_j+\mathcal{R}_j\right)-d_{i j} \geq \nonumber\\
&\sqrt{2} \operatorname{erf}^{-1}\left(1-2 \xi_{\text {coll }}^x\right) \chi,
\end{align}
where $\chi=\left\|\left[\begin{array}{c}
\big(\hat{\Sigma}_i^t\big)^{1 / 2}\left(\mathcal{A}_i+\mathcal{B}_i \mathcal{H}_i\right)^T E_x^{k^T} \Gamma^T \\
\Sigma_{\tilde{Z}_i}^{1 / 2}\left(\mathcal{K}_i+\mathcal{B}_i \mathcal{L}_i\right)^{\mathrm{T}} E_x^{k^T} \Gamma^T \\
\big(\tilde{\Sigma}_i^k\big)^{1 / 2} \Gamma^T \\
\big(\hat{\Sigma}_j^t\big)^{1 / 2}\left(\mathcal{A}_j+\mathcal{B}_j \mathcal{H}_j\right)^T E_x^{k^T} \Gamma^T \\
\Sigma_{\tilde{Z}_j}^{1 / 2}\left(\mathcal{K}_j+\mathcal{B}_j \mathcal{L}_j\right)^{\mathrm{T}} E_x^{k^T} \Gamma^T \\
\big(\tilde{\Sigma}_j^k\big)^{1 / 2} \Gamma^T
\end{array}\right]  \alpha_{ij}^k\right\|$ and $\alpha^k_{ij}$ is computed along the positions of the CAVs generated from the previous iteration based on the characteristics of the receding horizon optimization, which is similar to the successive linearization technique and can achieve sufficient accuracy

Then, we consider the relaxation of the control input chance constraint in Equation \ref{controlc} of each CAV. The feasible set $\mathcal{C}_u$ of the control input is denoted as the intersection of $q$ half-planes in $\mathbb{R}^{n_u}$, i.e., $\mathcal{C}_u=\bigcap_{q}\left\{u: \beta_q^{\top} u \leq b_q\right\} \subset \mathbb{R}^{n_u}$, where $\beta_q \in \mathbb{R}^{n_u}$ are constant vectors and $b_q \in \mathbb{R}$ are bounds for the control variables. Similar to the transformation of the state chance constraint and by leveraging the property of subadditivity of probabilities, the control input chance constraints can be transformed into 
\begin{equation}\label{controlc1}
\beta_q^T E_u^k \bar{U}_i-b_q \!\leq \!\sqrt{2} \operatorname{erf}^{-1}\!\left(2 \xi_{q, \!\text { fail }}^u\!\!-\!1\right)\!\left\|\!\!\left[\!\!\!\begin{array}{l}
\left(\hat{\Sigma}_i^t\right)^{1 / 2} \mathcal{H}_i^T E_u^{k^T} \\
\left(\Sigma_{\tilde{Z}_i}\right)^{1 / 2} \mathcal{L}_i^T E_u^{k^T}
\end{array}\!\!\!\right]\!\! \beta_q\right\|,
\end{equation}
where $\sum_{q=1}^2 \xi_{q, \text { fail }}^u \leq \xi_{\text {fail }}^u$ with $q=1, 2$ and $E_u^k \triangleq\left[\begin{array}{lll}0_{n_u, k n_u} & \mathrm{I}_{n_u} & 0_{n_u,(M-k-1) n_u}\end{array}\right]$.

Since the jerk is directly related to the control input and is affected by the random variables it contains, we convert the jerk constraint into the constraints on the mean and covariance. The jerk of each vehicle over the entire horizon is expressed in a compact form
\begin{equation}
\mathrm{K}_i =\frac{D_{\text{acc}} U_i}{\tau},    
\end{equation}
where $D_{\text{acc}}$ is the difference matrix and $\tau$ denotes the discrete-time interval. Thus, we can obtain
\begin{equation} \label{K_meancov}
\mathbb{E}\big[\mathrm{K}_i\big]=\left\|\frac{D_{\text{acc}} \bar{U}_i}{\tau}\right\|, \quad \Sigma_i^{\mathrm{K}_i} = \frac{D_{\text{acc}} \Sigma_i^U D_{\text{acc}}^T}{\tau^2}. 
\end{equation}
Let $E_\kappa^k \triangleq\left[\begin{array}{lll}0_{n_u, k n_u} & \mathrm{I}_{n_u} & 0_{n_u,(M-k-1) n_u}\end{array}\right]$, so that $E_\kappa^k \mathrm{K}_i = \kappa_i^k$, and substituting Equations \ref{K_meancov} and \ref{U_meancov}, the constraint in Equation \ref{jerk} can be converted as follows
\begin{equation}\label{a_mean}
\mathbb{E}\big[\kappa_i^k\big]=\left\|\frac{E_\kappa^k D_{\text{acc}} \bar{U}_i}{\tau}\right\| \leq \kappa_{\max }, 
\end{equation}
\begin{equation}\label{a_cov}
\resizebox{1\hsize}{!}{$
\lambda_{\max }\!\big(\!\Sigma_i^{\kappa_i^k}\!\big)\!\! =\!\!\!\lambda_{\max }\!\!\left(\!\!\frac{E_\kappa^k \! D_{\text{acc}} \! (\mathcal{H}_i \!\hat{\Sigma}_i^t \!\mathcal{H}_i^T \!\!+\! \mathcal{L}_i \Sigma_{\tilde{Z}_i} \!\!\mathcal{L}_i^T \!)D_{\text{acc}}^TE_\kappa^{k^T} \!}{\tau^2}\!\right) \!\!\! \leq \!\! \sigma_{\max }$},
\end{equation}
where $\sigma_{\max }$ limits the maximum eigenvalue of the jerk covariance matrix,  Equation \ref{a_mean} is convex, and Equation \ref{a_cov} can be transformed into equivalent form through the matrix norm, similar to \cite{okamoto2018optimal}
\begin{equation}\label{amatrix}
\left\|  \left[\!\!\!\begin{array}{l}
\frac{1}{\tau}\big(\hat{\Sigma}_i^t\big)^{1 / 2} \mathcal{H}_i^T D_{\text{acc}}^T E_\kappa^{k^T} \\
\frac{1}{\tau}\big(\Sigma_{\tilde{Z}_i}\big)^{1 / 2} \mathcal{L}_i^T D_{\text{acc}}^T E_\kappa^{k^T}
\end{array}\!\!\!\right] \right\|  \leq \sqrt{\sigma_{\max }}. 
\end{equation}
Thus, we derive two tractable convex constraints in Equations \ref{a_mean} and \ref{amatrix}.

Subsequently, the IM tries to minimize the objective in Equation \ref{longequ} with respect to the decision parameters $\boldsymbol{\bar{U}}, \boldsymbol{\mathcal{H}}$ and $\boldsymbol{\mathcal{L}}$, subject to the constraints in Equations \ref{avoidc111}, \ref{controlc1}, \ref{a_mean} and \ref{amatrix}. It follows that the objective and all constraints of the robust trajectory optimization problem are transformed into convex forms, enabling the problem to be solved efficiently using advanced convex optimization tools.

\subsection{Context-Aware Status Update Scheduler}\label{sectionIIIB}
After trajectory optimization iteration, we design the context-aware update scheduler. The task of the update scheduler is to prioritize state updates for the most urgent vehicles based on the \emph{vehicle driving context}. To align with coordination efforts and address the second challenge, we first present the formulation of the status update scheduling problem. Then we present a context-aware update scheme that features the explicit and flexible update index.

\textit{1) Problem Formulation:}
The IM needs to collect up-to-date states of the CAVs for coordination. To further enhance coordination safety in bandwidth limited environments, we aim to prioritize the most urgent vehicles to transmit their states to the IM. Therefore, the status update scheduling problem is formulated to align with coordination efforts.
To improve trajectory planning performance, the objective function in the status update scheduling process is first defined as 
\begin{equation}\label{obfcom}
J(\boldsymbol{V})=\lim _{T \rightarrow \infty} \mathbb{E}\left[\sum_{t=0}^{T-1} \sum_{i=1}^N\left(x_i^t-x_i^{r}\right)^T \!\!W_i^t\left(x_i^t-x_i^{r}\right)\!\right],
\end{equation}
which penalizes the deviation between the planned trajectory and reference trajectory. $W_i^t$ is the time-varying risk weight matrix associated with the risk level around the CAV. Herein, we define as a high-risk level when the CAV $i$ is in the CA to exemplify the proposed solution. Further, the safety assessment of vehicles can be extended to integrate with perception and prediction modules, thus providing a more comprehensive assessment of vehicle risk levels.
As previously stated, the time superscript $t$ denotes the $t$-th optimization iteration. By substituting $x_i^t=x_i^t-\bar{x}_i^t+\bar{x}_i^t$, the objective function is transformed into:
\begin{equation}\label{obfcom1}
J(\boldsymbol{V})\!\!=\!\!\!\lim _{T \rightarrow \infty} \left[\sum_{t=0}^{T-1}\! \sum_{i=1}^N\! \left(\bar{x}_i^t\!-\!x_i^{r}\right)^T \! W_i^t\!\left(\bar{x}_i^t\!-\!x_i^{r}\right)\!+\operatorname{tr}\!\left(W_i^t \Sigma_i^t\right)\!\right],
\end{equation}
where $\bar{x}_i^t$ and $\Sigma_i^t$ denote the mean and covariance of the initial state within each optimization iteration.

Then, the status update scheduling decisions at time $t$ are denoted as a vector $\boldsymbol{V}^t=\left(V_1^t, V_2^t, \cdots, V_N^t\right)$, where $V_i^t=1$ indicates that the $i$-th CAV is scheduled at time $t$, otherwise $V_i^t=0$. Due to stochastic fading effects, the successful transmission of status packets is defined as instantaneous SNR (Signal-to-Noise Ratio) exceeding a threshold.
The corresponding channel state indicator is represented by $S_i^t$, and the probability of transmission success is denoted as $s_i^t$.
Thus, there is a successful state delivery at time $t$ if and only if $V_i^t S_i^t=1$.
Due to the lack of channel resource, at each time $t$ a maximum of $n$ vehicles can be scheduled to simultaneously access the available wireless channel to transmit their status updates.
In particular, the maximum transmission frequency of each CAV is $\rho$, i.e.,
\begin{equation}\label{frequency}
\lim _{T \rightarrow \infty} \frac{1}{T} \sum_{t=0}^{T-1} V_i^t \leq \rho, \quad  \forall i \in \mathbb{N}
\end{equation}
\begin{equation}\label{maxnum}
\sum_{i=1}^N V_i^t \leq n, \quad \forall t \in \mathbb{N}.
\end{equation}

One can use first control input of the previous iteration to compute the estimated state mean and uncertainty of each CAV at the current optimization iteration that are given by
\begin{equation}\label{mt}
\bar{x}_i^{t}=A_i^{t-1} \bar{x}_i^{t-1}+B_i^{t-1} \bar{u}_i^{t-1},
\end{equation}
\begin{equation}\label{ecovt}
\hat{\Sigma}_i^{t}=\!\left(A_i^{t-\!1}\!\!+\!B_i^{t-\!1} H_i^{t-\!1}\right)\! \hat{\Sigma}_i^{t-\!1}\!\!\left(A_i^{t-\!1}\!\!+\!B_i^{t-\!1} H_i^{t-\!1}\right)^T\!\!\!+K_i^{t} \Sigma_{\tilde{z}_i^{t}} K_i^{t^T},
\end{equation}
where $\bar{u}_i^{t-1}$ and $H_i^{t-1}$ denote the first feedforward and feedback gain of the control input within the $(t-1)$-th iteration. The estimated state mean and uncertainty continue to evolve based on propagation Equations \ref{mt}, \ref{ecovt} at each iteration. Meanwhile, following \cite{zheng2020urgency}, the IM takes the received up-to-date states as initial points for coordination in the next iteration. Therefore, the dynamic function of the state mean and uncertainty in the status update process is expressed as
\begin{equation}\label{meant}
\bar{x}_i^{t}=V_i^{t} S_i^{t}\left(\hat{x}_i^{t}-\bar{x}_i^{t}\right)+\bar{x}_i^{t},
\end{equation}
\begin{align}\label{covt1}
\Sigma_i^{t}=\left(1-V_i^{t} S_i^{t}\right)\hat{\Sigma}_i^{t}+\tilde{\Sigma}_i^{t},
\end{align}
where $\hat{x}_i^{t}$ is the filtered state obtained by each CAV at the beginning of the $t$-th iteration.


To enhance context-aware timeliness of status updates, we try to adaptively schedule the status update of each CAV. We can omit the constant term $x_i^{r}$ in Equation \ref{obfcom1} and take the average of the objective function. Therefore, the status update scheduling problem in our coordination framework is formulated as
\begin{align*}
\begin{gathered}
\mathscr{P}_2:\min _{V_i^t} \underset{T \rightarrow \infty}{\lim } \frac{1}{N T} \sum_{t=0}^{T-1} \sum_{i=1}^N \bar{x}_i^{t^T} \!W_i^t \bar{x}_i^t+ \operatorname{tr}\left(W_i^t \Sigma_i^t\right) \\
\text { s.t. } \lim _{T \rightarrow \infty} \frac{1}{T} \sum_{t=0}^{T-1} V_i^t \leq \rho, \quad  \forall i \in \mathbb{N}\\
\sum_{i=1}^N V_i^t \leq n, \quad \forall t \in \mathbb{N}
\end{gathered}
\end{align*}
where the dynamic function of the state mean $\bar{x}_i^t$ and covariance $\Sigma_i^t$ of each CAV are Equations \ref{meant} and \ref{covt1}. Through carefully designing scheduling decision $V_i^t$ at each time slot, we aim to improve the effectiveness of status updates.

\textit{2) Scheduling Scheme:} 
In order to satisfy the average transmission frequency constraint in Equation \ref{frequency}, the virtual queue $Y_i^t$ could be defined as
\begin{equation}
Y_i^{t+1}=\left[Y_i^t-\rho +V_i^{t} \right]^{+},
\end{equation}
where $[x]^{+}=\max (0, x)$, The length of the virtual queue serves as an indicator of historical transmission budget usage: when the CAV performs a status update, the virtual queue $Y_i^t$ increases by $1-\rho$; otherwise, the virtual queue decreases by $\rho$. As long as the virtual queue maintains rate stability, the frequency constraint in Equation \ref{frequency} is naturally guaranteed \cite{neely2022stochastic}.

In order to obtain a feasible scheduling scheme, the Lyapunov function is defined as $I_t=\sum_{i=1}^N \theta_i (Y_i^t)^2$, where $\theta_i$ is a positive real number to trade off performance and queue stability. Then the Lyapunov drift is
\begin{equation}
\Delta I_t=\mathbb{E}\left[I_{t+1}-I_t \mid \bar{\boldsymbol{x}}^t, \hat{\boldsymbol{\Sigma}}^t, \boldsymbol{W}^{t}, \boldsymbol{Y}^t\right].
\end{equation}
At the same time, let the real penalty be the cost at the $t$-th iteration, i.e., $f_t=\sum_{i=1}^N \bar{x}_i^{t^T} \! W_i^{t} \bar{x}_i^{t}+\operatorname{tr}\left(W_i^{t} \Sigma_i^{t}\right)$. 
According to Lyapunov optimization theory \cite{neely2022stochastic}, the problem of reducing the objective function while enforcing the frequency constraints can be simplified to reduce the expected sum of the Lynapunov drift and penalty. 
Given estimated state mean $\bar{\boldsymbol{x}}^t$, covariance $\hat{\boldsymbol{\Sigma}}^t$, context-aware weight $\boldsymbol{W}^{t}$, as well as virtual queue length $\boldsymbol{Y}^t$ at the $t$-th iteration, the upper bound of the expected drift-plus-penalty is derived as
\begin{equation}
\resizebox{1\hsize}{!}{$
\begin{aligned}\label{driftpluspenalty}
&\Delta I_t \!+ \mathbb{E} \!\left[f_t \!\mid \! \bar{\boldsymbol{x}}^t \!, \hat{\boldsymbol{\Sigma}}^t \!\!, \boldsymbol{W}^{t}\!\!, \boldsymbol{Y}^t\!\right] \!\!	\leq \! \sum_{i=1}^N \!\! \Big\{\!2  \theta  Y_i^t \!+ \! s \operatorname{tr}\!\big[W_i^{t}\big(\hat{x}_i^{t} \hat{x}_i^{t^T}\!\!\!\!-\!\bar{x}_i^{t} \bar{x}_i^{t^T}\big)\\
&-\!W_i^{t}\hat{\Sigma}_i^{t}\big] \!\Big\} \! V_i^{t} \!\!+\! \sum_{i=1}^N \!\!\Big[\! \operatorname{tr}\!\big(W_i^{t} \hat{\Sigma}_i^{t} \!+ W_i^{t} \tilde{\Sigma}_i^{t}\! + W_i^{t} \bar{x}_i^{t} \bar{x}_i^{t^T} \big)\!\! -\! 2\theta \rho Y_i^t \!+ \!\theta  \Big]. 
\end{aligned}$}
\end{equation}

Therefore, minimizing the right-hand side of the above drift-plus-penalty inequality is equivalent to obtaining the scheduling problem in the following form
\begin{align}\label{scheme}
\min _{V_i^t} \sum_{i=1}^N \! \Big\{2  \theta  Y_i^t + & s \operatorname{tr}\!\big[W_i^{t} \big(\hat{x}_i^{t} \hat{x}_i^{t^T}\!\!\!-\bar{x}_i^{t} \bar{x}_i^{t^T}\big)\!-W_i^{t}\hat{\Sigma}_i^{t}\big] \Big\}V_i^{t} \nonumber \\
&\text { s.t. } \sum_{i=1}^N V_i^{t} \leq n .
\end{align}
We can find that the objective function in Equation \ref{scheme} comprises a linear combination of scheduling decisions $V_i^{t}$. Therefore, under the constraint of limited wireless resources, the update index of CAV $i$ at the $t$-th optimization iteration is defined as
\begin{equation}\label{index}
\begin{aligned}
\lambda_i^{t}=2  \theta  Y_i^t +  s \operatorname{tr}\!\big[W_i^{t}\big(\hat{x}_i^{t} \hat{x}_i^{t^T}\!\!-\bar{x}_i^{t} \bar{x}_i^{t^T}\big)-W_i^{t}\hat{\Sigma}_i^{t}\big].
\end{aligned}
\end{equation}

The solution to Equation \ref{scheme} is to prioritize scheduling the $n$ urgent CAVs with the smallest update indices $\lambda_i^{t}$. Thus, we can find that the context-aware scheme grants a higher probability of update to the CAV that has a larger state uncertainty $\hat{\Sigma}_i^{t}$, a greater risk weight $W_i^{t}$, a more accurate filtered state $\hat{x}_i^{t}$ or a shorter virtual queue length $Y_i^t$. These factors can be summarized as critical state conditions and potential surrounding risks, i.e., \emph{vehicle’s driving context}, which reflects the urgency of driving. In other words, the update index measures the necessity of status transmission by fully assessing the urgency of vehicle driving. Additionally, the derived update index can be tightly coupled with the trajectory planner due to its receding horizon form. Finally, applying context-aware update decisions can reduce potential driving risks for the most urgent $n$ CAVs.


\subsection{Proposed Robust Intersection Coordination Framework}\label{sectionIIIC}

In outline, the proposed robust and comprehensive intersection coordination framework is summarized in Algorithm \ref{Algorithm 1}.
Next, we analyze the computational complexity of the algorithm in the two phase.
First, for the robust trajectory planning phase, the number of entries to be computed under the truncated affine disturbance feedback control policy is $\mathcal{O}\left(N M n_x n_u\right)$~\cite{balci2021covariance}. The resulting convex formulation is solved through the Mosek solver with interior-point solution. The key to the complexity of interior-point methods for cone programming lies in the set of linear equations involved, whose scale is primarily determined by the dimensionality of decision variables and second-order cone constraints for collision avoidance \cite{andersen2011interior}. It follows that the approximate computational complexity of the cooperative trajectory planner scales with $O_p = \mathcal{O}\Big((N M n_x n_u+M^2 {n_x}^2({n_u}^2+1))^3\Big)$. 
Second, the computational complexity of the adaptive state update phase mainly lies in calculating the update index, with a complexity of $O_u=\mathcal{O}\left(N{n_x}^3\right)$.
Finally, the overall complexity is $O_p+O_u$. 

\begin{algorithm}

	\small
	\caption{The Robust Intersection Coordination Framework}
	\label{Algorithm 1}
	\LinesNumbered
	\KwIn{Horizon $M$, Probability thresholds $\xi_{\text {coll }}^x$ and $\xi_{\text {fail }}^u$, Process noise matrix $G$, Measurement and noise matrix $C$, $D$, Maximum transmission frequency $\rho$, Number of sub-channels $n$, Transmission success probability $s$;}
	Initialization: $t$:= $0$; $\big(\hat{x}_i^{0}, \hat{\Sigma}_i^0, \tilde{\Sigma}_i^{0^{-}}\big)$: Initial estimation state of each CAV, $i \in \mathcal{V}$\;
	Randomly generate CAVs entry times $t_i^e$ into the CCZ\;
	\Repeat{\emph{all CAVs reach their destinations}}{
            calculate the update index by Equation \ref{index} to determine the update decision $\boldsymbol{V}^t$ at time $t$\;
            Sub-channel authorization, upload states to the IM according to the update decision $\boldsymbol{V}^t$\;
            $\bar{\boldsymbol{X}}^t$, $\bar{\boldsymbol{U}}^t$:= get nominal trajectory based on previous control sequence\;
		$(\boldsymbol{A}^k)_{k=t}^{t+M-1}\!, (\boldsymbol{B}^k)_{k=t}^{t+M-1}\!, (\boldsymbol{r}^k)_{k=t}^{t+M-1}$:= linearize dynamics using Equation \ref{linear}, \ref{residual} around nominal trajectory\;
		$\boldsymbol{\mathcal{A}}$, $\boldsymbol{\mathcal{B}}$, $\boldsymbol{\mathcal{R}}$, $\boldsymbol{\mathcal{K}}$:= construct the block matrices\;
            Check and adapt the inter-vehicle collision avoidance coupling based on current positions $\boldsymbol{p}^t$\;
		$\big(\bar{\boldsymbol{U}}, \boldsymbol{\mathcal{H}}, \boldsymbol{\mathcal{L}}\big)$:= calculate robust coordination 
            problem by Equations \ref{longequ}, \ref{avoidc111}, \ref{controlc1}, \ref{a_mean} and \ref{amatrix}\;
		Obtain the control input $\boldsymbol{U}$ by Equation \ref{feedbackforward} and execute the first command\;
		Remove vehicles upon arrival at destinations\;
		$t$:= $t+1$\;
	}
\end{algorithm}

\section{Simulation Results}\label{sectionV}
In this section, we validate the performance of our coordination framework through extensive simulations. The considered intersection scenario is shown in Fig. \ref{Fig01}. The road geometry is set as \cite{luo2023real}. 
The random entry times of vehicles in the CCZ follow the widely adopted Poisson distribution \cite{guo2017spatial}. The arrival rate is set at 1.2 vehicles/lane/s to simulate the high traffic level at the intersection, based on traffic classification from previous work \cite{luo2023real}. Furthermore, vehicle flow consists of 25\% right-turning vehicles, with left-turning and straight-through vehicles each accounting for 37.5\%, based on the vehicle patterns established in \cite{donglin2024TITS}.
We set the probability thresholds $\xi_{\text {coll}}^x$ = 0.1 and $\xi_{\text { fail }}^u$ = 0.05 to ensure that CAVs cross intersections with a suitable level of safety and to facilitate a clear comparison and evaluation of the effects of various factors on coordination performance~\cite{knaup2023safe}.
The measurement noise matrix is set as $D = \mathrm{diag}(0.4,0.2,\pi/150,0.1)$ based on the localization error levels recorded in real-world experiments in \cite{li2018high,bai2022time}. The discrete-time interval $\tau$ and wheelbase parameters $L_w$ are set to 0.1 s and 2.7 m, respectively, based on the configurations used in \cite{luo2023real}.
The transmission success probability $s$ (i.e., transmission reliability) and system transmission rate (i.e., Tx rate [message/s]) are set at 0.95 and 10, respectively, according to the recommendations in \cite{abdel20205g}. Vehicles in the CA are more prone to collisions than others due to narrow inter-vehicle spacing and frequent interactions. We set the risk weight $W$ as $\mathrm{diag}(10, 10, 10, 10)$ when vehicles enter the CA, and as $\mathrm{diag}(1, 1, 1, 1)$ otherwise.

All experiments are conducted in MATLAB R2019b using YALMIP (version R20210331) for optimization modeling and MOSEK (version 9.3.20) as the solver. Specifically, using YALMIP, we define the optimization variables and then directly employ them to symbolically express the objective function and constraints based on the derived formulations. MOSEK is invoked through YALMIP’s solver interface, using the interior-point method with default parameter settings~\cite{aps2019mosek}. No additional parameter tuning was performed.
The entire simulation design is divided into three main components: 1) The scene and vehicle configuration component sets the information of the roads and vehicles, where each vehicle is designed as a ``struct'' to define and update relevant information. 2) The coordination computation component is configured for computing robust cooperative trajectories and state updating priorities, which is regarded as fulfilling the role of IM. 3) The environment refresh component is configured for vehicle movement and scene refresh visualization. Furthermore, after the scene and vehicle configuration, the entire coordination process iteratively performs coordination computation and environment refresh using a receding horizon fashion.
The detailed parameters are summarized in Table~\ref{simupara}.




\setlength{\tabcolsep}{1.45mm}{
	\begin{table}[h]
		\vspace{-.5em} 
		\caption{Simulation Parameters}
		\centering
		\label{simupara}
		\begin{tabular}{cc}		
			\toprule[1pt]
			Parameters & Values \\
			\midrule
			Size of CA   & 20 m $\times$ 20 m \\
			Size of CCZ  & 100 m $\times$ 100 m \\
                left/right-turning radius & 15 m / 5 m \\
			Vehicle size $L_{\mathrm{car}} / W_{\mathrm{car}}$  &4 m / 2 m \\
			Maximum velocity $v_{max}    $   & 20 m/s    \\
			Maximum acceleration $a$          &$[-5,5]$ m/s$^2$ \\  
			Maximum steering angle $\delta$     &$[-0.78,0.78]$ rad  \\	
			Minimum safety distance $d_{ij}  $   & 4 m   \\
                time slot size $\tau$  \quad      & 100 ms     \\
			Time horizon $M$          & 20    \\ 
		    State weight matrix $Q$          &$\operatorname{diag}(10,10,1,1)$ \\
                Terminal state weight matrix $Q^M$ &$\operatorname{diag}(50,50,1,1)$ \\
			Input weight matrix $R$          &$\operatorname{diag}(20,20)$ \\
			Maximum transmission frequency $\rho    $   & 0.95    \\
			Initial state covariance $\hat{\Sigma}_i^0$  & $\mathrm{diag}(0.1,0.05,\pi/180,0.02)$  \\ 
			Prior estimation error covariance $\tilde{\Sigma}_i^{0^-}$  & $\mathrm{diag}(0.02,0.01,\pi/360,0.02)$ \\
			Process noise matrix $G$        & $\mathrm{diag}(0.03,0.02,\pi/180,0.1)$ \\
			Measurement matrix $C$        & $I_4$     \\     
			\bottomrule[1pt]
		\end{tabular}
\end{table}}

\subsection{Performance of the Proposed Coordination Framework}\label{sectionVB}

In this subsection, we conducted extensive comparative experiments in uncertain and bandwidth-limited environments to validate the performance of our proposed framework. The collision probability serves as the safety metric and the total passing time serves as the efficiency metric. 
we measure the passing time per run through 200 simulated runs of intersection crossing. As the differences in collision outcomes across different vehicle numbers are minor, especially when the number of vehicles is small, we performed an increased number of simulations (i.e., 500 runs) for the statistics on collision probability.
We compare the safety and efficiency of our coordination framework with state-of-the-art coordination methods. There are three benchmarks considered:

\noindent $\bullet$ Space-time resource searching (STRS) \cite{zhang2021trajectory}: It successively searches the XYT three-dimensional resource blocks to obtain the trajectories and passing order.

\noindent $\bullet$ Re-planning  STRS (Re-STRS)\cite{chen2022re}: It introduces the variable safety redundancy and re-planning strategy by considering error accumulation in the search for resource blocks.

\noindent $\bullet$ Stochastic model predictive control (SMPC): It applies the standard SMPC approach to coordinate vehicles, which optimizes feed-forward control inputs while utilizing a fixed stabilizing feedback gain\cite{arcari2023stochastic}.

\begin{figure}[htbp] 
	\centering
	\begin{subfigure}{0.49\columnwidth}
		\centering
		\includegraphics[width=\textwidth,trim=118 65 105 53,clip]{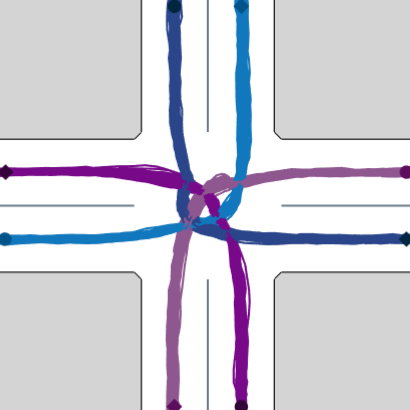}
		\vspace{-1em}
		\caption{SMPC}	
	\end{subfigure}
	\centering
	\begin{subfigure}{0.49\columnwidth}
		\centering
		\includegraphics[width=\textwidth,trim=118 65 105 53,clip]{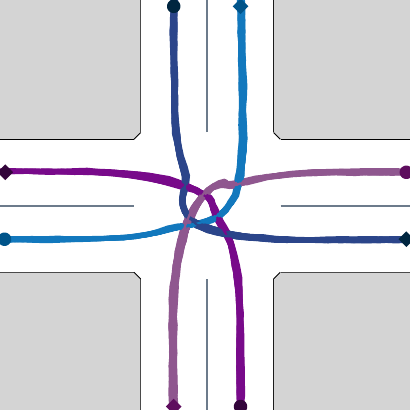}
		\vspace{-1em}
		\caption{Proposed} 
	\end{subfigure}
	\caption{Visualization of 100 simulated vehicle centroid trajectories. (a) SMPC. (b) Proposed. The output trajectories of the proposed method exhibit the smallest spread, which can reduce the potential collisions in multi-vehicle interactions.}
	\label{robust}
\end{figure}

Initially, we focus on the performance of robust cooperative trajectory planner, for which we provide sufficient sub-channel resources. 
The robustness of the proposed framework and the SMPC approach was tested through 100 Monte Carlo simulations of intersection crossing. Following the setup in\cite{luo2023real}, four left-turning CAVs are initialized to enter the intersection. The circle and diamond markers represent the start and end points of a vehicle, respectively. As demonstrated in Fig.~\ref{robust}, the proposed framework achieves the smallest spread in the output trajectories. The collision instance count also shows a marked reduction. The proposed framework records 4 collision instances compared to the 14 instances in the SMPC approach, highlighting greater robustness to random disturbances. This result can be explained by the coupled mean-covariance collision avoidance constraint and the fact that our method actively steers the covariance of the vehicle positions.

\begin{figure}
	\centering
	\includegraphics[width=1.0\linewidth]{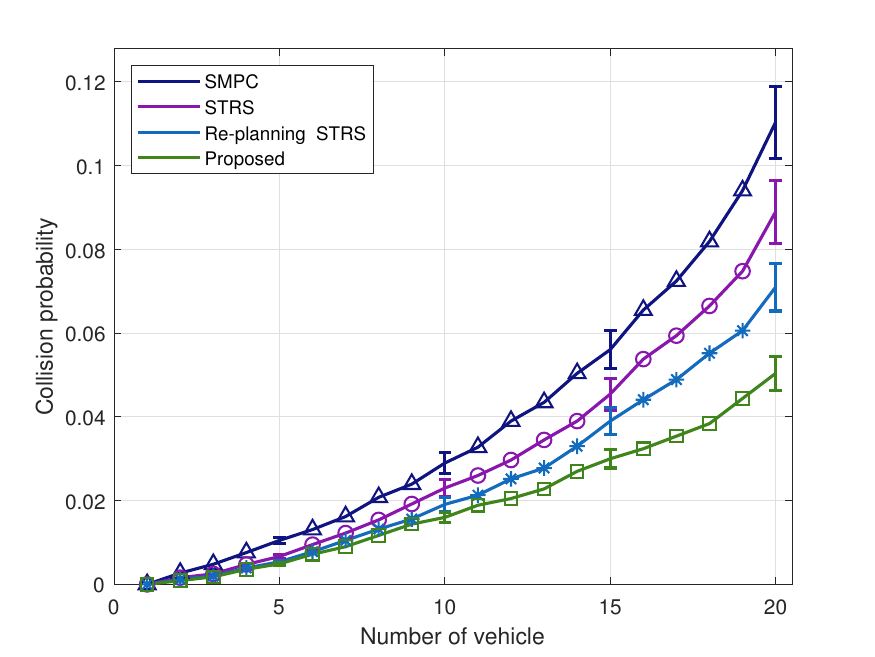}\\
	\vspace{-0.5em}
	\caption{Collision probability of different benchmarks.}\label{th1}
	\vspace{-1em}
\end{figure}

The collision probability of the four coordination solutions is demonstrated in Fig. \ref{th1}. It is obvious that a greater number of CAVs could increase the collision probability, due to the increased frequency and extent of interactions between CAVs. The collision probability of SMPC is higher than that of STRS, Re-STRS, and our framework. The Re-STRS framework exhibits a lower collision probability than STRS due to its variable safety redundancy which adjusts based on the accumulation of trajectory deviation in noisy conditions. The collision probability of our framework is almost the same as that of the Re-STRS framework when the number of vehicles is less than 9, and it is markedly lower when the number of vehicles exceeds 9. This is because our framework dynamically and precisely characterizes uncertainty and its evolution, providing more accurate error capture than the variable safety redundancy of the Re-STRS framework. This leads to improved safety benefits, especially as the number of vehicles increases.

\begin{figure}
	\centering
	\includegraphics[width=1.0\linewidth]{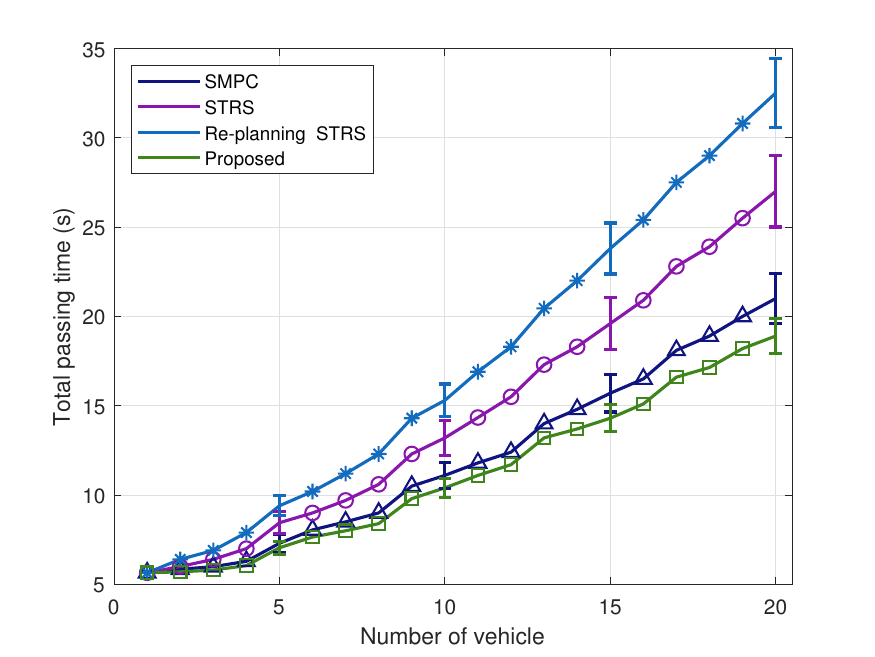}\\
	\vspace{-0.5em}
	\caption{Total passing time of different benchmarks.}\label{throught}
	\vspace{0em}
\end{figure}

Fig. \ref{throught} compares the total passing time of four coordination solutions. It is obvious that the total passing time of our framework is marginally less than SMPC and significantly less than the STRS and Re-STRS frameworks when the number of vehicles in the intersection increases. The total passing time of the Re-STRS framework is greater than that of the STRS framework since variable safety redundancy significantly reduces the set of searchable resource blocks. The efficiency advantage of our solution corresponds to the fact that the proposed framework simultaneously computes vehicle maneuvers within the future horizon, whereas STRS and Re-STRS allocate resources to vehicles sequentially. This more efficient use of road spatial resources results in higher traffic throughput.

\begin{table*}[!t]
	\captionsetup{font={small}}
	\caption{\centering{\scshape Safety and Efficiency Comparison of Coordination Solutions At Different Numbers of CAVs With Sufficient Sub-channels, CP: Collision Probability (\%); TPT: Total Passing Time (s)}}
	\label{tab3}
	\renewcommand\tabcolsep{3pt} 
	\centering
	\begin{tabular}{cc@{\hspace{10pt}}c@{\hspace{10pt}}cc@{\hspace{10pt}}cc@{\hspace{10pt}}cc@{\hspace{5pt}}}
		\toprule
		\multirow{2}{*}{Num CAV} & \multicolumn{2}{c}{Proposed} & \multicolumn{2}{c}{Re-STRS\cite{chen2022re}} & \multicolumn{2}{c}{STRS\cite{zhang2021trajectory}} & \multicolumn{2}{c}{SMPC\cite{arcari2023stochastic}}  \\	
		& CP  & TPT   & CP & TPT  & CP & TPT & CP & TPT \\
		\midrule
		5     &0.4±0.03	          &\textbf{7.05±0.36}	&0.4±0.03  &9.40±0.55	&0.6±0.05   &8.45±0.63	&1.0±0.07  &7.34±0.51 \\
		10	  &\textbf{1.6±0.12}  &\textbf{10.42±0.54}	&2.0±0.16  &15.33±0.92	&2.4±0.20   &13.20±0.99	&3.0±0.26  &11.13±0.72 \\
		15	  &\textbf{3.0±0.22}  &\textbf{14.31±0.76}	&4.0±0.32  &23.75±1.42	&4.6±0.38   &19.66±1.45	&5.6±0.46  &15.71±1.03 \\
		20	  &\textbf{5.0±0.40}  &\textbf{18.90±0.98}	&7.0±0.58  &32.21±1.95	&9.0±0.76   &27.07±1.99	&11.0±0.88 &21.09±1.38 \\
		\bottomrule		
	\end{tabular}
    \vspace{-1em}
\end{table*}

Table \ref{tab3} shows the collision probability and total passing time of the four solutions under different numbers of CAVs. For total passing time, our framework provides average improvements of 32.25\%, 21.63\%, and 6.44\% in coordination efficiency compared to the Re-STRS, STRS, and SMPC methods, respectively. 
While the total passing time of our framework is slightly lower than that of the SMPC method, it outperforms SMPC in terms of collision probability. Our framework with 0.05 is significantly better than the SMPC with 0.110, which means $54.5\%$ advantage in coordination safety. Therefore, the proposed framework not only significantly reduces collision probability but also maintains a comparable advantage in throughput. Moreover, our framework exhibits the smallest performance fluctuations in terms of collision probability and total passing time, demonstrating stable coordination robustness.

\begin{table}[!t]
	\captionsetup{font={small}}
	\caption{\centering{\scshape Safety and Efficiency Comparison of Coordination Solutions At Different Numbers of Sub-channels With a Fixed 20 CAVs, CP: Collision Probability (\%); TPT: Total Passing Time}(s)}
	\label{tab4}
	\renewcommand\tabcolsep{0.5pt}
	\centering
	\begin{tabular}{c@{\hspace{5pt}}c@{\hspace{5pt}}c@{\hspace{8pt}}c@{\hspace{5pt}}c@{\hspace{5pt}}}
		\toprule
		\multirow{2}{*}{Num channel } & \multicolumn{2}{c}{Proposed} & \multicolumn{2}{c}{Re-STRS\cite{chen2022re}}  \\	
		& CP  & TPT   & CP & TPT   \\
		\midrule
		20		&\textbf{5.0±0.40}	&\textbf{18.90±0.98}	&7.0±0.58	&32.21±1.95  \\
		15      &\textbf{5.6±0.42}  &\textbf{19.26±1.00}	&8.0±0.65	&32.98±1.98	 \\
		10		&\textbf{7.0±0.55}	&\textbf{19.92±1.05}	&10.2±0.82	&34.04±2.06   \\
		5		&\textbf{9.4±0.76}	&\textbf{20.87±1.11}	&13.8±0.97	&35.23±2.18	  \\
		\bottomrule		
	\end{tabular}
    \vspace{-1em}
\end{table}

Previous results highlight the performance advantages of our approach in terms of both safety and efficiency, while also showing the variations in the results under varying traffic conditions. A concise and comprehensive explanation is provided here.
The variation in collision probability can be attributed to the following two factors. An increased number of vehicles introduces a higher frequency and extent of interactions between CAVs. Furthermore, vehicle congestion limits the available passing space, posing challenges in finding optimal solutions within the limited space, which in turn reduces trajectory flexibility. Thus, these factors increase the potential for collisions resulting from vehicle interactions. 
The total passing time of our coordination framework is approximately proportional to the number of vehicles, which indicates that the passing time required for each vehicle is essentially consistent. This is because vehicles are able to pass through the intersection at a steady speed without significant deceleration, highlighting the efficiency advantages of our coordination framework.

Subsequently, we compare the performance of the proposed coordination framework and Re-STRS under insufficient sub-channel resources. STRS and SMPC methods cannot operate in bandwidth-limited environments because they lack the corresponding state update process design. Insufficient sub-channel resources make it unsustainable to ensure that all vehicles update their states in a timely manner at every moment. This affects the evolution of vehicle state uncertainty, theoretically impacting the real-time safety and efficiency of coordination. The number of CAVs is kept constant at 20. Table \ref{tab4} presents the results of the two methods under different numbers of sub-channels. It can be observed that as the sub-channels decrease, the collision probability and total passing time for Re-STRS increase significantly more rapidly than in our framework. 
This could be explained by the effective incorporation of robust trajectory control and adaptive state updates in our framework. 
At 5 sub-channels, our framework reduces the collision probability and total passing time by 31.88\% and 40.76\%, respectively, compared to Re-STRS. This shows that our framework maintains significant coordination safety and efficiency advantages even in practical bandwidth-limited environments.

\subsection{Evaluation of the Context-aware Status Update Scheduler}\label{sectionVC}

In this subsection, we further evaluate the effectiveness of the proposed status update scheduling scheme to investigate the effect of the state update design on coordination safety. Two scheduling benchmarks are compared in the simulation: 

\noindent $\bullet$ Round robin: Vehicles are scheduled to update sequentially in a fixed order.

\noindent $\bullet$ AOI based: The top $n$ vehicles with the highest value of $\Delta_i^t(\Delta_i^t+1)$ are scheduled, where $\Delta_i^t$ denotes the AOI of the $i$-th vehicle at time $t$. This scheme is shown to be asymptotically AoI-optimal in \cite{jiang2019unified}.

\begin{figure*}[htbp]
	\centering
	\begin{subfigure}{1\columnwidth}
		\centering
		\includegraphics[width=\textwidth]{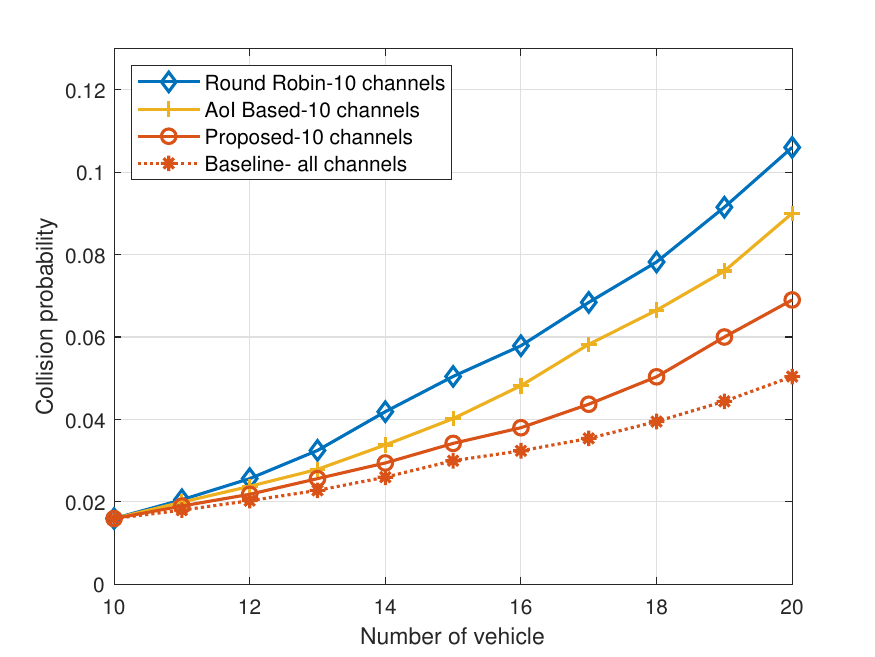}
		\vspace{-1.5em}
		\caption{Collision probability of different scheduling schemes.}	\label{diffcomms}
	\end{subfigure}
	\begin{subfigure}{1\columnwidth}
		\centering
		\includegraphics[width=\textwidth]{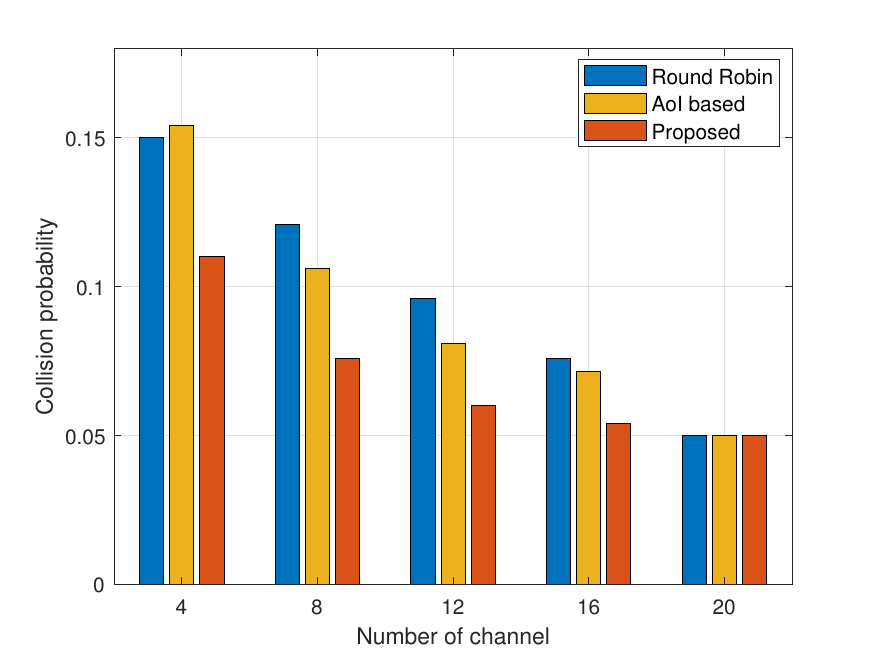}
		\vspace{-1.5em}
		\caption{Collision probability with different channel resources.} 
		\label{numchann}
	\end{subfigure}
	\vspace{-0.5em}
	\caption{Impact of bandwidth resource utilization on coordination safety}
	\vspace{-1.5em}
	\label{impactcomm}
\end{figure*}

Fig. \ref{impactcomm} presents the collision probability of these schemes under different bandwidth resource conditions. An important observation in Fig. \ref{diffcomms} is that our proposed context-aware scheme significantly outperforms other scheduling schemes in terms of collision probability. There are $n$ = 10 sub-channels available for CAVs to send states to the IM. It is obvious that as the number of vehicles increases, the collision probability of three scheduling schemes with 10 sub-channels significantly exceeds and grows faster than the corresponding scheme with sufficient sub-channels (i.e., baseline). 
The round robin scheme has the highest collision probability. Our context-aware scheme reduces the collision probability by around $25\%$ compared to the AoI-based scheme. Moreover, our scheme results in a significant reduction in collision probability over round-robin and AoI-based schemes, especially when the number of vehicles is large. This indicates that the context-aware and rational use of wireless resources offers a more pronounced improvement in coordination safety performance.

We further investigated how bandwidth resources affect the collision probability. The number of vehicles is kept constant at 20. 
Fig. \ref{numchann} demonstrates the collision probability of three schemes versus different channel resources. It can be observed that as the available channel resources for status transmissions, i.e., $n$, increase, the collision probability continuously decreases. This reduction is due to the corresponding decrease in the state uncertainty of the vehicles, which implies the trade-off between resource utilization and coordination safety. 
Compared to other schemes, our scheme exhibits a more pronounced reduction in collision probability when there are few channel resources. Furthermore, as the number of sub-channels exceeds 12, the collision probabilities in our scheme decrease and become nearly identical to those under conditions of ample channel resources. This indicates that our scheme could require fewer bandwidth resources while still providing a comparable level of safety performance.

\subsection{Discussion: Robustness of the Proposed Framework}\label{sectionVD}

In this subsection, we discuss the robustness of the proposed framework through two finer-grained studies. We highlight the influence of different levels of measurement noise, failure probabilities, and transmission success probabilities on coordination reliability. 

 \begin{figure}
	\centering
	\includegraphics[width=1.0\linewidth]{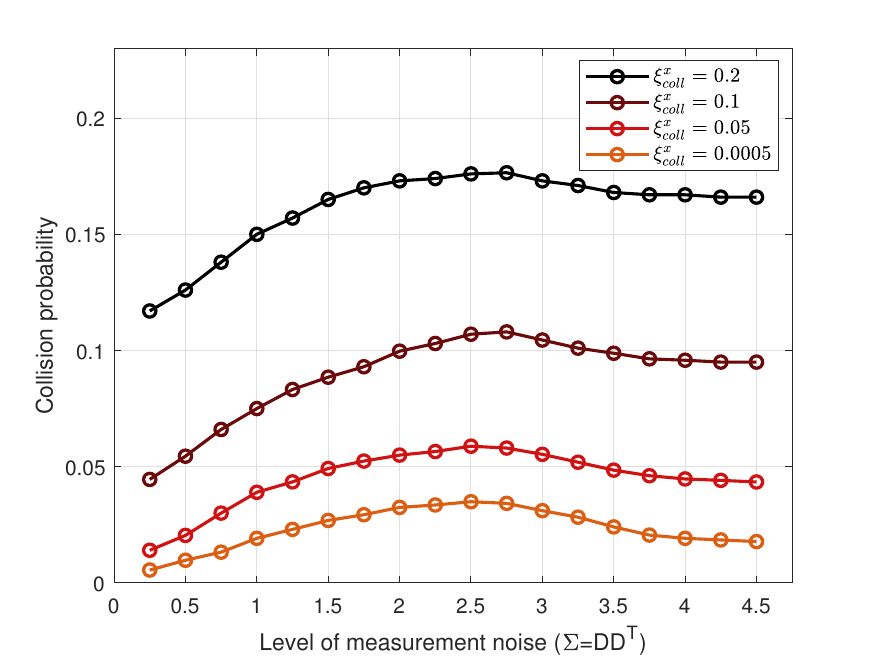}\\
	\vspace{-0.5em}
	\caption{Safety performance of different collision probabilities and levels of measurement noise.}\label{measurnoise}
	\vspace{-1em}
\end{figure}

Fig. \ref{measurnoise} shows the collision probability under different failure probabilities and levels of measurement noise. The number of vehicles and sub-channels remains at 20 and 10, respectively. 
From the figure, as the failure probability keeps decreasing, it ensures greater certainty in vehicle interactions and hence leads to a smaller collision probability.
Meanwhile, when the measurement noise is low, the collision probability increases accordingly as the measurement noise level rises. However, upon exceeding a measurement noise level of approximately $2.75\Sigma$, a decrease in collision probability is observed. This corresponds to the fact that the filtered state tends to rely more on prior estimation rather than on the most recent measurement when the measurement noise is considerable. This indicates that the proposed framework is robust to higher levels of noise.

\begin{figure}
	\centering
	\includegraphics[width=1.0\linewidth]{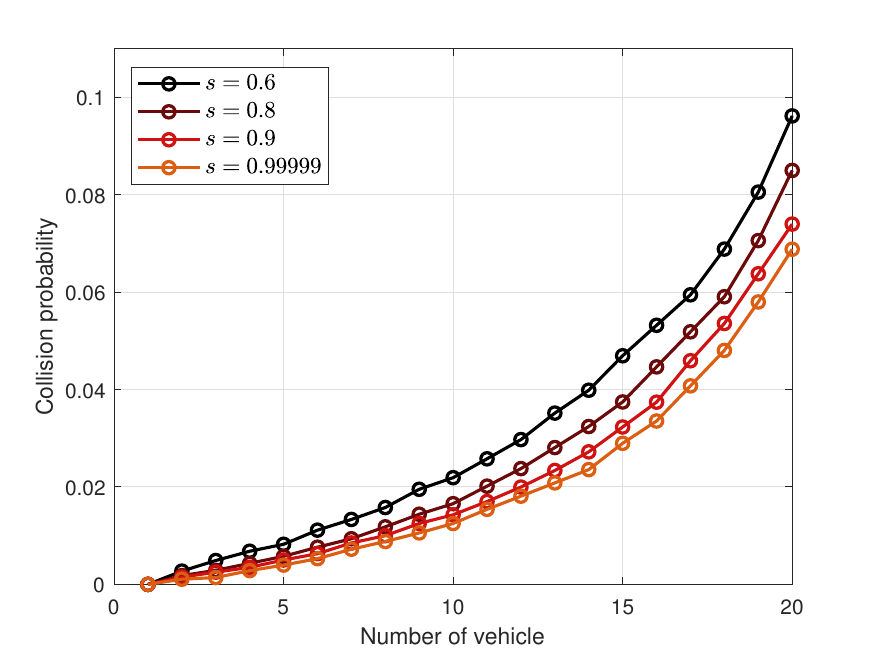}\\
	\vspace{-0.5em}
	\caption{Safety performance of different transmission success probabilities.}\label{failureproba}
	\vspace{-1em}
\end{figure}

In Fig. \ref{failureproba}, we present how the transmission success probability influences the collision probability. The number of sub-channels is kept constant at 10. It can be observed that the collision probability increases as the transmission success probability decreases. This is because the vehicle state covariance increases accordingly as the packet losses increase. In this process, packet losses cause the covariance to evolve in a predictive manner without any sensor feedback.

\begin{table}[h] 
	\captionsetup{font={small}}
	\caption{\centering{\scshape Safety and Efficiency Comparison of the Proposed Method under Different Transmission Success Probability Settings with 20 CAVs, CP: Collision Probability (\%); TPT: Total Passing Time(s)}}
        \setlength{\tabcolsep}{30pt}
        \centering
	\begin{tabular}{c@{\hspace{20pt}}c@{\hspace{25pt}}c}
		\toprule
            \multirow{1}{*}{Probability $P_s$ } & CP & TPT   \\	
		\midrule
		Fixed 0.95	     &7.0	 &19.92	   \\
		  0.905 - 0.995    &7.2    &20.16	 \\
		\bottomrule
	\end{tabular}\label{Probability}
\end{table}
    Table \ref{Probability} further illustrates the impact of dynamically varying transmission success probabilities on coordination safety and efficiency.
    The Rayleigh fading channel is adopted and the transmission process is influenced by path loss and small-scale fading \cite{weber2010overview}.
    A transmission is considered successful if the received SNR exceeds a predefined threshold $\gamma$. The transmission success probability is given by $\exp \big(-\gamma \frac{N_0}{P_{\mathrm{tx}}} d^\eta\big)$, where $\eta$, $\gamma$, $N_0$, and $P_{\mathrm{tx}}$ denote the path loss exponent, SNR threshold, noise power, and transmission power, and $d$ denotes the distance between the vehicle and the IM. These parameters are set to 3, 16dB, -99dBm, and -18dBm to emulate a realistic urban canyon environment \cite{haneda2016frequency}. Accordingly, the transmission success probability is computed to lie within the range of 0.905 to 0.995. 
    Compared to the fixed transmission success probability of 0.95, the dynamically varying success probabilities in the range of 0.905–0.995 led to a 2.9\% increase in collision probability and a 1.2\% increase in total passing time.
    Furthermore, as shown in Fig. 9, when 20 CAVs are involved, the collision probabilities for fixed transmission success probabilities of 0.9 and 1 are 6.8\% and 7.4\%, respectively. It can be observed that the collision probability for the varying range of 0.905–0.995 falls within this interval.
    The dynamically varying transmission success probabilities resulting from position changes and fading indeed introduce additional randomness into the state update process. However, this variability can be averaged over the interval boundaries, and the resulting performance differences remain relatively minor.
    
\begin{table}[h] 
	\captionsetup{font={small}}
	\caption{\centering{\scshape Collisions (/500) Comparison of the Proposed Method at Different Collision Probability Thresholds $\xi_{\text {coll }}^x$ } }
	\centering
        \footnotesize
	\begin{tabular}{c@{\hspace{2pt}}c@{\hspace{5pt}}c@{\hspace{5pt}}c}
		\toprule
            \multirow{1}{*}{Num CAV } & $\xi_{\text {coll }}^x\!\!\!=\!\!10^{-2}$ & $\xi_{\text {coll }}^x\!\!\!=\!\!10^{-3}$ & $\xi_{\text {coll }}^x\!\!\!=\!\!10^{-4}$  \\	
		\midrule
		5  & 1   & \textbf{0}  & 0 \\
		10 & 2   & \textbf{0}  & 0 \\
		15 & 4  & \textbf{0}  & 2 \\
		20 & 4  & \textbf{2}  & 3 \\
		\bottomrule
	\end{tabular}\label{coll}
\end{table}
    Table \ref{coll} shows the collision results of our framework under very low collision probability thresholds, ranging from 0.01 to 0.0001. We adopt the batch optimization method proposed in our previous work \cite{luo2023real}, which effectively manages vehicle flows to achieve more orderly and safer coordination. Here, the batch size is set to 2. The sub-channel resources are sufficient. Since the difference in the output of the collision probability is minimal, we measure the number of collisions as the number of vehicles increases through 500 simulated runs of intersection crossing.
    From the Table \ref{coll}, as the collision probability threshold decreases, the number of collisions also decreases, indicating an improvement in safety.
    An interesting observation is that when the number of vehicles exceeds 15, the number of collisions at a probability threshold of $10^{-4}$ is higher than that at $10^{-3}$. It reveals that excessively low probability thresholds result in unstable collision avoidance performance, which does not yield consistent safety improvements. This is because the excessively low probability threshold in the chance constraints imposes overly stringent safety requirements. Specifically, vehicles maintain larger and redundant safety margins, which significantly limit the available passage space and flexibility of trajectories, potentially leading to infeasibility of the problem, especially in complex or dynamic interactions. Therefore, setting the threshold to $10^{-3}$ appropriately balances strict safety requirements with coordination efficiency and feasibility.

\section{Conclusion}\label{sectionVI}
This paper studies the intersection control of CAVs in practical uncertainty environments. The proposed framework successively determines the robust trajectories and state update priorities for CAVs by integrating a robust cooperative trajectory planner and a context-aware status update scheduler, which copes with various sources of state uncertainty, including motion uncertainty,
sensor measurement noise, and transmission imperfections. 
In particular, the robust cooperative trajectory planner directly controls the evolution of state uncertainty in a receding horizon fashion and can tightly couple with the flexible update index derived by the context-aware status update scheduler.
Simulation results show that the proposed framework outperforms existing strategies in terms of robustness, safety, and efficiency. Furthermore, our framework demonstrates superior effectiveness in utilizing bandwidth resources compared to benchmark scheduling schemes. It clearly shows that our framework is robust to high levels of uncertainty and transmission limitations arising from the practical system.

Future research directions involve an extension of the presented framework to a hybrid coordination structure (centralized and distributed). We will investigate semantics-aware communication strategies to reduce the amount of uninformative data exchanges and support scalable vehicle collaborations~\cite{luo2025TCOM, luo2025TIT}. Moreover, we will implement more efficient robust coordination algorithms, including sequential reformulation methods and offline feedback matrix computation. 

\begin{table}[h]
        \begin{centering}
        \caption[Acronym used in this chapter]{Acronym used in this paper.\label{Acronym}}
		\par\end{centering}
	\noindent\resizebox{1.0\columnwidth}{!}{%
		\begin{centering}
			\begin{tabular}{|>{\centering}m{0.25\columnwidth}|>{\raggedright}m{0.86\columnwidth}|}
				\hline 
				\textbf{Acronym} & \textbf{\qquad \qquad \qquad \qquad \qquad \qquad Definition}\tabularnewline
				\hline 
				\hline 
				CAVs  &  Connected and autonomous vehicles.\tabularnewline				
				\hline 
                    ITS  & Intelligent transportation systems. \tabularnewline
				\hline
                    GPS  & Global position system. \tabularnewline
				\hline
                     3GPP & 3rd Generation Partnership Project. \tabularnewline
				\hline
                    NR & New Radio. \tabularnewline
				\hline
                    OBU & Vehicle onboard unit. \tabularnewline
				\hline
				C-V2X  & Cellular Vehicle-to-everything. \tabularnewline
				\hline
				V2I & Vehicle-to-infrastructure. \tabularnewline
				\hline
				CCZ  & Centralized control zone. \tabularnewline
				\hline
				CA  & Conflict area. \tabularnewline
				\hline
                    IM & Intersection manager. \tabularnewline
				\hline
                    MAC & Multi-access computing. \tabularnewline
				\hline
				SNR &  Signal-to-noise ratio. \tabularnewline
				\hline
                    STRS  & Space-time resource searching method. \tabularnewline
				\hline
				Re-STRS  & Re-planning Space-time resource searching method. \tabularnewline
				\hline
				SMPC &  Stochastic model predictive control method. \tabularnewline
				\hline
                    CP  & Collision probability. \tabularnewline
				\hline
				TPT &  Total passing time. \tabularnewline
				\hline
                    AOI &  Age of information. \tabularnewline
				\hline
			\end{tabular}
			\par\end{centering}
	}
\end{table}

\bibliographystyle{IEEEtran}
\bibliography{reference}

\begin{thebibliography}{10}
\providecommand{\url}[1]{#1}
\csname url@samestyle\endcsname
\providecommand{\newblock}{\relax}
\providecommand{\bibinfo}[2]{#2}
\providecommand{\BIBentrySTDinterwordspacing}{\spaceskip=0pt\relax}
\providecommand{\BIBentryALTinterwordstretchfactor}{4}
\providecommand{\BIBentryALTinterwordspacing}{\spaceskip=\fontdimen2\font plus
\BIBentryALTinterwordstretchfactor\fontdimen3\font minus \fontdimen4\font\relax}
\providecommand{\BIBforeignlanguage}[2]{{%
\expandafter\ifx\csname l@#1\endcsname\relax
\typeout{** WARNING: IEEEtran.bst: No hyphenation pattern has been}%
\typeout{** loaded for the language `#1'. Using the pattern for}%
\typeout{** the default language instead.}%
\else
\language=\csname l@#1\endcsname
\fi
#2}}
\providecommand{\BIBdecl}{\relax}
\BIBdecl

\bibitem{garcia2021tutorial}
M.~H.~C. Garcia, A.~Molina-Galan, M.~Boban, J.~Gozalvez, B.~Coll-Perales, T.~{\c{S}}ahin, and A.~Kousaridas, ``A tutorial on 5g nr v2x communications,'' \emph{IEEE Communications Surveys \& Tutorials}, vol.~23, no.~3, pp. 1972--2026, 2021.

\bibitem{rios2016survey}
J.~Rios-Torres and A.~A. Malikopoulos, ``A survey on the coordination of connected and automated vehicles at intersections and merging at highway on-ramps,'' \emph{IEEE Transactions on Intelligent Transportation Systems}, vol.~18, no.~5, pp. 1066--1077, 2016.

\bibitem{luo2023computationally}
J.~Luo, T.~Zhang, and Q.~Zhang, ``A computationally efficient bi-level coordination framework for cavs at unsignalized intersections,'' \emph{IEEE Transactions on Vehicular Technology}, 2023.

\bibitem{khayatian2020survey}
M.~Khayatian, M.~Mehrabian, E.~Andert, R.~Dedinsky, S.~Choudhary, Y.~Lou, and A.~Shirvastava, ``A survey on intersection management of connected autonomous vehicles,'' \emph{ACM Transactions on Cyber-Physical Systems}, vol.~4, no.~4, pp. 1--27, 2020.

\bibitem{namazi2019intelligent}
E.~Namazi, J.~Li, and C.~Lu, ``Intelligent intersection management systems considering autonomous vehicles: A systematic literature review,'' \emph{IEEE Access}, vol.~7, pp. 91\,946--91\,965, 2019.

\bibitem{mitrovic2019combined}
N.~Mitrovic, I.~Dakic, and A.~Stevanovic, ``Combined alternate-direction lane assignment and reservation-based intersection control,'' \emph{IEEE Transactions on Intelligent Transportation Systems}, vol.~21, no.~4, pp. 1779--1789, 2019.

\bibitem{meng2017analysis}
Y.~Meng, L.~Li, F.-Y. Wang, K.~Li, and Z.~Li, ``Analysis of cooperative driving strategies for nonsignalized intersections,'' \emph{IEEE Transactions on Vehicular Technology}, vol.~67, no.~4, pp. 2900--2911, 2017.

\bibitem{luo2023real}
J.~Luo, T.~Zhang, R.~Hao, D.~Li, C.~Chen, Z.~Na, and Q.~Zhang, ``Real-time cooperative vehicle coordination at unsignalized road intersections,'' \emph{IEEE Transactions on Intelligent Transportation Systems}, 2023.

\bibitem{guan2020centralized}
Y.~Guan, Y.~Ren, S.~E. Li, Q.~Sun, L.~Luo, and K.~Li, ``Centralized cooperation for connected and automated vehicles at intersections by proximal policy optimization,'' \emph{IEEE Transactions on Vehicular Technology}, vol.~69, no.~11, pp. 12\,597--12\,608, 2020.

\bibitem{liu2024cooperative}
J.~Liu, P.~Hang, X.~Na, C.~Huang, and J.~Sun, ``Cooperative decision-making for cavs at unsignalized intersections: A marl approach with attention and hierarchical game priors,'' \emph{IEEE Transactions on Intelligent Transportation Systems}, 2024.

\bibitem{pei2019cooperative}
H.~Pei, S.~Feng, Y.~Zhang, and D.~Yao, ``A cooperative driving strategy for merging at on-ramps based on dynamic programming,'' \emph{IEEE Transactions on Vehicular Technology}, vol.~68, no.~12, pp. 11\,646--11\,656, 2019.

\bibitem{zhang2021trajectory}
Y.~Zhang, R.~Hao, T.~Zhang, X.~Chang, Z.~Xie, and Q.~Zhang, ``A trajectory optimization-based intersection coordination framework for cooperative autonomous vehicles,'' \emph{IEEE Transactions on Intelligent Transportation Systems}, vol.~23, no.~9, pp. 14\,674--14\,688, 2021.

\bibitem{donglin2024TITS}
D.~Li, T.~Zhang, J.~Luo, T.~Liang, B.~Cao, X.~Wu, and Q.~Zhang, ``A tightly coupled bi-level coordination framework for cavs at road intersections,'' \emph{IEEE Transactions on Intelligent Transportation Systems}, vol.~25, no.~7, pp. 7832--7847, 2024.

\bibitem{rios2016automated}
J.~Rios-Torres and A.~A. Malikopoulos, ``Automated and cooperative vehicle merging at highway on-ramps,'' \emph{IEEE Transactions on Intelligent Transportation Systems}, vol.~18, no.~4, pp. 780--789, 2016.

\bibitem{dey2015review}
K.~C. Dey, L.~Yan, X.~Wang, Y.~Wang, H.~Shen, M.~Chowdhury, L.~Yu, C.~Qiu, and V.~Soundararaj, ``A review of communication, driver characteristics, and controls aspects of cooperative adaptive cruise control ({CACC}),'' \emph{IEEE Transactions on Intelligent Transportation Systems}, vol.~17, no.~2, pp. 491--509, 2015.

\bibitem{bai2022time}
X.~Bai, W.~Wen, and L.-T. Hsu, ``Time-correlated window-carrier-phase-aided gnss positioning using factor graph optimization for urban positioning,'' \emph{IEEE Transactions on Aerospace and Electronic Systems}, vol.~58, no.~4, pp. 3370--3384, 2022.

\bibitem{li2018high}
T.~Li, H.~Zhang, Z.~Gao, Q.~Chen, and X.~Niu, ``High-accuracy positioning in urban environments using single-frequency multi-{GNSS RTK/MEMS-IMU} integration,'' \emph{Remote sensing}, vol.~10, no.~2, p. 205, 2018.

\bibitem{chalaki2021priority}
B.~Chalaki and A.~A. Malikopoulos, ``A priority-aware replanning and resequencing framework for coordination of connected and automated vehicles,'' \emph{IEEE Control Systems Letters}, vol.~6, pp. 1772--1777, 2021.

\bibitem{vitale2022autonomous}
C.~Vitale, P.~Kolios, and G.~Ellinas, ``Autonomous intersection crossing with vehicle location uncertainty,'' \emph{IEEE Transactions on Intelligent Transportation Systems}, vol.~23, no.~10, pp. 17\,546--17\,561, 2022.

\bibitem{vitale2022optimizing}
C.~\vspace{0mm}Vitale, P.~\vspace{0mm}Kolios, and G.~\vspace{0mm}Ellinas, ``Optimizing vehicle re-ordering events in coordinated autonomous intersection crossings under cavs' location uncertainty,'' \emph{IEEE Transactions on Intelligent Vehicles}, vol.~8, no.~5, pp. 3473--3488, 2022.

\bibitem{pan2023hierarchical}
X.~Pan, B.~Chen, L.~Dai, S.~Timotheou, and S.~A. Evangelou, ``A hierarchical robust control strategy for decentralized signal-free intersection management,'' \emph{IEEE Transactions on Control Systems Technology}, vol.~31, no.~5, pp. 2011--2026, 2023.

\bibitem{chen2022re}
C.~Chen, J.~Luo, T.~Liang, and T.~Zhang, ``Re-planning optimization of cooperative vehicle coordination at road intersections,'' in \emph{2022 IEEE 95th Vehicular Technology Conference:(VTC2022-Spring)}, 2022, pp. 1--6.

\bibitem{okamoto2018optimal}
K.~Okamoto, M.~Goldshtein, and P.~Tsiotras, ``Optimal covariance control for stochastic systems under chance constraints,'' \emph{IEEE Control Systems Letters}, vol.~2, no.~2, pp. 266--271, 2018.

\bibitem{zheng2020urgency}
X.~Zheng, S.~Zhou, and Z.~Niu, ``Urgency of information for context-aware timely status updates in remote control systems,'' \emph{IEEE Transactions on Wireless Communications}, vol.~19, no.~11, pp. 7237--7250, 2020.

\bibitem{li2024toward}
A.~Li, S.~Wu, S.~Meng, R.~Lu, S.~Sun, and Q.~Zhang, ``Toward goal-oriented semantic communications: New metrics, framework, and open challenges,'' \emph{IEEE Wireless Communications}, 2024.

\bibitem{luo2025TCOM}
J.~Luo and N.~Pappas, ``Semantic-aware remote estimation of multiple {Markov} sources under constraints,'' \emph{IEEE Trans. Commun.}, May 2025, early access.

\bibitem{zheng2024c}
J.~Zheng, B.~Wang, and C.~Li, ``A c-v2x mode 4 and 802.11 p based resource selection scheme for intra-platoon message delivery,'' \emph{IEEE Internet of Things Journal}, 2024.

\bibitem{wang2023inp}
B.~Wang, J.~Zheng, N.~Mitton, and C.~Li, ``Inp-crs: an intra-platoon cooperative resource selection scheme for c-v2x networks,'' \emph{IEEE Communications Letters}, vol.~27, no.~11, pp. 3118--3122, 2023.

\bibitem{luo2025TIT}
J.~Luo and N.~Pappas, ``On the cost of consecutive estimation error: Significance-aware non-linear aging,'' \emph{IEEE Transactions on Information Theory}, pp. 1--1, July 2025, early access.

\bibitem{nazari2018remote}
M.~A. Nazari, T.~Charalambous, J.~Sj{\"o}berg, and H.~Wymeersch, ``Remote control of automated vehicles over unreliable channels,'' in \emph{2018 IEEE Wireless Communications and Networking Conference (WCNC)}, 2018, pp. 1--6.

\bibitem{nazari2018impact}
M.~A. Nazari, A.~Ozcelikkale, M.~Zanon, T.~Charalambous, J.~Sj{\"o}berg, and H.~Wymeersch, ``Impact of communication frequency on remote control of automated vehicles,'' in \emph{2018 IEEE 29th Annual International Symposium on Personal, Indoor and Mobile Radio Communications (PIMRC)}, 2018, pp. 96--100.

\bibitem{chohan2019robust}
N.~Chohan, M.~A. Nazari, H.~Wymeersch, and T.~Charalambous, ``Robust trajectory planning of autonomous vehicles at intersections with communication impairments,'' in \emph{2019 57th Annual Allerton Conference on Communication, Control, and Computing (Allerton)}, 2019, pp. 832--839.

\bibitem{bai2023context}
H.~Bai, H.~Li, W.~Dou, and Y.~Wang, ``Context-aware timely status updates for trajectory control with limited communication resources,'' in \emph{2023 IEEE 97th Vehicular Technology Conference (VTC2023-Spring)}, 2023, pp. 1--6.

\bibitem{paden2016survey}
B.~Paden, M.~{\v{C}}{\'a}p, S.~Z. Yong, D.~Yershov, and E.~Frazzoli, ``A survey of motion planning and control techniques for self-driving urban vehicles,'' \emph{IEEE Transactions on intelligent vehicles}, vol.~1, no.~1, pp. 33--55, 2016.

\bibitem{zheng2024cs}
D.~Zheng, J.~Ridderhof, Z.~Zhang, P.~Tsiotras, and A.-A. Agha-Mohammadi, ``Cs-brm: A probabilistic roadmap for consistent belief space planning with reachability guarantees,'' \emph{IEEE Transactions on Robotics}, 2024.

\bibitem{zhu2019chance}
H.~Zhu and J.~Alonso-Mora, ``Chance-constrained collision avoidance for mavs in dynamic environments,'' \emph{IEEE Robotics and Automation Letters}, vol.~4, no.~2, pp. 776--783, 2019.

\bibitem{ridderhof2020chance}
J.~Ridderhof, K.~Okamoto, and P.~Tsiotras, ``Chance constrained covariance control for linear stochastic systems with output feedback,'' in \emph{2020 59th IEEE Conference on Decision and Control (CDC)}, 2020, pp. 1758--1763.

\bibitem{balci2021covariance}
I.~M. Balci and E.~Bakolas, ``Covariance control of discrete-time gaussian linear systems using affine disturbance feedback control policies,'' in \emph{2021 60th IEEE Conference on Decision and Control (CDC)}.\hskip 1em plus 0.5em minus 0.4em\relax IEEE, 2021, pp. 2324--2329.

\bibitem{aastrom2012introduction}
K.~J. {\AA}str{\"o}m, \emph{Introduction to stochastic control theory}.\hskip 1em plus 0.5em minus 0.4em\relax Courier Corporation, 2012.

\bibitem{neely2022stochastic}
M.~Neely, \emph{Stochastic network optimization with application to communication and queueing systems}.\hskip 1em plus 0.5em minus 0.4em\relax Springer Nature, 2022.

\bibitem{andersen2011interior}
M.~Andersen, J.~Dahl, Z.~Liu, L.~Vandenberghe, S.~Sra, S.~Nowozin, and S.~Wright, ``Interior-point methods for large-scale cone programming,'' \emph{Optimization for machine learning}, vol. 5583, 2011.

\bibitem{guo2017spatial}
J.~Guo, Y.~Zhang, X.~Chen, S.~Yousefi, C.~Guo, and Y.~Wang, ``Spatial stochastic vehicle traffic modeling for vanets,'' \emph{IEEE Transactions on Intelligent Transportation Systems}, vol.~19, no.~2, pp. 416--425, 2017.

\bibitem{knaup2023safe}
J.~Knaup, K.~Okamoto, and P.~Tsiotras, ``Safe high-performance autonomous off-road driving using covariance steering stochastic model predictive control,'' \emph{IEEE Transactions on Control Systems Technology}, vol.~31, no.~5, pp. 2066--2081, 2023.

\bibitem{abdel20205g}
S.~A. Abdel~Hakeem, A.~A. Hady, and H.~Kim, ``5g-v2x: Standardization, architecture, use cases, network-slicing, and edge-computing,'' \emph{Wireless Networks}, vol.~26, no.~8, pp. 6015--6041, 2020.

\bibitem{aps2019mosek}
M.~ApS, ``Mosek optimization toolbox for matlab,'' \emph{User’s Guide and Reference Manual, Version}, vol.~4, no.~1, 2019.

\bibitem{arcari2023stochastic}
E.~Arcari, A.~Iannelli, A.~Carron, and M.~N. Zeilinger, ``Stochastic mpc with robustness to bounded parametric uncertainty,'' \emph{IEEE Transactions on Automatic Control}, vol.~68, no.~12, pp. 7601--7615, 2023.

\bibitem{jiang2019unified}
Z.~Jiang, S.~Zhou, Z.~Niu, and C.~Yu, ``A unified sampling and scheduling approach for status update in multiaccess wireless networks,'' in \emph{IEEE INFOCOM 2019-IEEE Conference on Computer Communications}.\hskip 1em plus 0.5em minus 0.4em\relax IEEE, 2019, pp. 208--216.

\bibitem{weber2010overview}
S.~Weber, J.~G. Andrews, and N.~Jindal, ``An overview of the transmission capacity of wireless networks,'' \emph{IEEE Transactions on Communications}, vol.~58, no.~12, pp. 3593--3604, 2010.

\bibitem{haneda2016frequency}
K.~Haneda, N.~Omaki, T.~Imai, L.~Raschkowski, M.~Peter, and A.~Roivainen, ``Frequency-agile pathloss models for urban street canyons,'' \emph{IEEE Transactions on Antennas and Propagation}, vol.~64, no.~5, pp. 1941--1951, 2016.

\end{thebibliography}

\end{document}